\documentclass[11pt]{article}  
\usepackage[left=1in, right=1in, top=1in, bottom=1in]{geometry}
\usepackage{physics,hyperref,amsmath,amssymb,graphics,mathalfa,mathtools,colortbl,lipsum}
\usepackage[demo]{graphicx}
\usepackage{caption}
\usepackage{subcaption}
\numberwithin{equation}{section}
\usepackage{silence}
\usepackage{pythonhighlight}
\usepackage[]{hyperref}
\usepackage{array}
\makeatletter
\g@addto@macro\bfseries{\boldmath}
\makeatother

\gdef\@journal{jhep}

\newif\ifnotoc\notocfalse
\newif\ifemailadd\emailaddfalse
\newif\iftoccontinuous\toccontinuousfalse
\newif\ifnatbibsort\natbibsorttrue

\relax
\usepackage{longtable,caption}
\newcolumntype{L}[1]{>{\raggedright\let\newline\\\arraybackslash\hspace{0pt}}m{#1}}
\newcolumntype{C}[1]{>{\centering\let\newline\\\arraybackslash\hspace{0pt}}m{#1}}
\newcolumntype{R}[1]{>{\raggedleft\let\newline\\\arraybackslash\hspace{0pt}}m{#1}}
\usepackage[framemethod=TikZ]{mdframed}
\mdfdefinestyle{sid}{%
    linecolor=black,
    outerlinewidth=1.0pt,
    roundcorner=7pt,
    innerrightmargin=15pt,
    innerleftmargin=15pt,
    backgroundcolor=gray!30
		}	
\definecolor{Gray}{gray}{0.8}
\definecolor{MyBlue}{rgb}{0.0,0.0,0.9}
\definecolor{MyRed}{rgb}{0.0,0.9,0.0}
\colorlet{Bluee}{MyBlue!6}
\colorlet{Redd}{MyRed!1}
\DeclareMathAlphabet{\mathcal}{OMS}{cmsy}{m}{n} 

\usepackage{blindtext}
\title{\textbf{Exploring topological entanglement through Dehn surgery}}
\date{}
\author{\textbf{\emph{Aditya Dwivedi}}$^a$, \textbf{\emph{Siddharth Dwivedi}}$^b$, \textbf{\emph{Vivek Kumar Singh}}$^c$,\\
\textbf{\emph{Pichai Ramadevi}}$^d$, \textbf{\emph{Bhabani Prasad Mandal}}$^a$,  \thanks{Email: aditya.dwivedi13@bhu.ac.in, siddharth.dwivedi@curaj.ac.in, vks2024@nyu.edu, ramadevi@phy.iitb.ac.in, bhabani@bhu.ac.in}\\ \\ 
     $^a$ Department of Physics, Institute of Science, Banaras Hindu University,
Varanasi, 221005, India\\ $^b$ Department of Physics, Central University of Rajasthan,
Ajmer, 305817, India\\ $^c$ Center for Quantum and Topological Systems (CQTS),
NYUAD Research Institute,\\
New York University Abu Dhabi, PO Box 129188, Abu Dhabi, UAE\\$^d$ Department of Physics, Indian Institute of Technology Bombay,
Powai, Mumbai, 400076, India  }

\begin{document}
\maketitle
\begin{abstract}
We compute the $\text{PSL}(2,\mathbb{C})$ Chern-Simons partition function of a closed 3-manifold obtained from Dehn fillings of the link complement $\mathbf S^3\backslash {\mathcal{L}}$, where $\mathcal{L}=\mathcal{K}\# H$ is the connected sum of the knot $\mathcal {K}$ with the  Hopf link $H$. Motivated by our earlier work on topological entanglement and the reduced density matrix $\sigma$ for such link complements, we wanted to determine a choice of Dehn filling so that the trace of the matrix $\sigma$ becomes equal to the  $\text{PSL}(2,\mathbb{C})$ partition function of the closed 3-manifold. We use the \texttt{SnapPy} program and numerical techniques to show this equivalence up to the leading order. We have given explicit results for all hyperbolic knots $\mathcal{K}$ up to six crossings.
\end{abstract}
\tableofcontents
\newpage
\section{Introduction} \label{sec1}
Chern-Simons theory, a prominent example of a topological quantum field theory, has been extensively studied in both physics and mathematics. This theory provides a wide-ranging applications across various domains \cite{witten1989}\cite{lab1999}\cite{heinonen1998}\cite{murthy2003}\cite{marino2005}. While the Chern-Simons theory with compact gauge groups has been in the limelight due to its deep connection with knot theory \cite{witten1989}, the interest in exploring the Chern-Simons theory with complex gauge groups has surged in recent years. The latter provides novel applications in gravity theories and defining previously undiscovered invariants for $3$-manifolds \cite{witten1991}\cite{gukov2005}\cite{Balasubramanian:2018por}\cite{Dimofte:2016pua}\cite{Duan:2022ryd}\cite{Freed:2022yae}\cite{Garoufalidis:2023ipa}.

In this work, we analyze the partition function of the closed manifolds that result from the Dehn fillings of link complement manifolds in the complex Chern-Simons theory with $\text{PSL}(2,\mathbb{C})$ gauge group. The motivation behind this study is due to our recent work \cite{ad} on the multi-boundary entanglement in the context of compact Chern-Simons theory with $\text{SO}(3)$ gauge group.

In \cite{ad}, we studied the entanglement features of the pure states $\ket{\mathcal{K} \# H}$ where $\mathcal {K} \# H$ denotes a connected sum of a knot $\mathcal {K}$ and the Hopf link $H$. These states are prepared by computing the Chern-Simons path integral on the link complement $\mathbf S^3  \backslash (\mathcal{K} \# H)$  based on the SO(3) gauge group. Numerically the inner product $\bra{\mathcal{K} \# H}\ket{\mathcal{K} \# H}$ is equal to the trace of the unnormalized reduced density matrix associated with this state which we denoted as $\text{Tr}[\sigma(\mathcal{K} \# H)]$ in \cite{ad}. Topologically the inner product $\bra{\mathcal{K} \# H}\ket{\mathcal{K} \# H}$ can be viewed as taking two copies of $\mathbf S^3 \backslash \mathcal (\mathcal {K} \# H)$ and gluing them along their respective oppositely oriented boundaries resulting in a closed 3-manifold $M_{\mathcal{K}}$. If we denote `ZSO' to be the SO(3) Chern-Simons partition function, then we must have:
\begin{equation}
\text{Tr}[\sigma(\mathcal{K} \# H)] = \text{ZSO}_k(M_{\mathcal{K}}) ~,
\end{equation}
where $k$ is the level of the SO(3) Chern-Simons theory. Although the exact topology of $M_{\mathcal{K}}$ was not known to us, we discovered in \cite{ad} that the asymptotic or the large $k$ limit of this partition function gives the volume of $\mathbf S^3  \backslash \mathcal{K}$. The precise statement is the following \cite{ad}:
\begin{equation} \label{cor1}
\lim_{k \to \infty} \frac{\ln \text{ZSO}_k(M_{\mathcal{K}})}{k} = \frac{\text{Vol}(\mathbf S^3 \backslash \mathcal {K})}{2\pi} ~.
\end{equation}

In the present article, we try to understand the density matrix $\sigma(\mathcal{K} \# H)$ or equivalently the manifold $M_{\mathcal{K}}$ from a different perspective. In particular, we want to see whether the manifold $M_{\mathcal{K}}$ can be obtained from the Dehn surgery of $\mathbf S^3  \backslash (\mathcal{K} \# H)$. We focus on Dehn fillings of the link complement $\mathbf S^3  \backslash (\mathcal{K} \# H)$ to get a plausible quantitative 
understanding of $M_{\mathcal {K}}$. To do this, we compute the $\text{PSL}(2,\mathbb{C})$ Chern-Simons partition functions of the closed 3-manifolds resulting from the Dehn fillings of $\mathbf S^3  \backslash (\mathcal{K} \# H)$ using the machinery of the state integral model \cite{gang, hikami1, hikami2}. Exploiting the connection between the $\text{PSL}(2,\mathbb{C})$ and SO(3) Chern-Simons theories\footnote{From the conjecture 5 of \cite{gang}, it is clear that the asymptotic expansion of  $\text{PSL}(2,\mathbb{C})$ partition function numerically coincides with the SO(3) partition function.}, we look for the Dehn fillings for which the asymptotic expansion of the $\text{PSL}(2,\mathbb{C})$ partition function satisfies the limit similar to that of \eqref{cor1}. We shall use the numerical calculations in the state integral approach to obtain the asymptotic limit of the partition functions.

The paper is organized as follows: In section \ref{sec2}, we discuss the preliminaries related to our setup. We include a  brief review of the complex Chern-Simons theory and the state integral approach in this section. In section \ref{sec3}, we present the numerical results. In section \ref{sec4}, we summarize the results and also highlight some future directions.
Appendix \ref{triangulation} and appendix \ref{appB}  contain the details of the ideal triangulation of the manifold and volume formula after $(p,q)$ Dehn filling respectively. These details will provide the necessary background for the examples discussed in section \ref{sec3}.
\section{Preliminaries} \label{sec2}
\subsection{State integral partition function in complex Chern-Simons theory} 
The action of the complex Chern-Simons theory is defined on a 3-manifold $M$ as follows:
\begin{equation}
\mathcal{S}  =  \frac{t}{8 \pi}
\int_{M} \Tr
\left( \mathcal{A} \wedge d \mathcal{A} + \frac{2}{3}\,\mathcal{A} \wedge \mathcal{A} \wedge \mathcal{A} \right) \label{csaction} 
 +\;  \frac{\Tilde{t} }{8 \pi} \int_{M} \Tr \Big( \Tilde{\mathcal{A}}  \wedge d \Tilde{\mathcal{A}}
+ \frac{2}{3}\Tilde{\mathcal{A}} \wedge \Tilde{\mathcal{A}} \wedge \Tilde{\mathcal{A}} \Big)\,,
\end{equation}
where the gauge field $\mathcal{A}$ has a complex nature and the coupling constants $t$ and $\Tilde{t}$ can be expressed as $t=(k+\theta)$ and $\Tilde{t}=(k-\theta)$\footnote{It is important to note that, although $t$ and $\Tilde{t}$ are not necessarily complex conjugates of each other, the parameter $k$ must be an integer, and the parameter $\theta$ must be either real or purely imaginary to ensure the consistency and unitarity of the quantum theory, as discussed in \cite{witten1991}.}. The partition function of the Chern-Simons theory based on a complex gauge group $\mathbb G_C$ can be computed using the Feynman path integral:
\begin{equation}
Z_{\hbar}^{\mathbb G_C}(M) = \int D\mathcal{A}\, e^{i\mathcal{S}}\label{int} ~,
\end{equation}
where $D\mathcal{A}$ denotes the integration over all possible gauge field configurations and $\hbar$ is inversely proportional to $k$. When the manifold $M$ has the boundary $\partial M$, the above partition function gives a state in the Hilbert space associated with $\partial M$. Out of various manifolds, the most interesting ones are those where $M$ is the link complement manifold of the type $\mathbf S^3\backslash \mathcal L$ where $\mathcal L$ is a link.  In the following, we will discuss the necessary details of the  $\text{PSL}(2,\mathbb{C})$ Chern-Simons partition function obtained using the state integral model \cite{gang}. 

Let $M=\mathbf S^3\backslash \mathcal L$ be the link-complement manifold where $\mathcal L$ is a link made of $n$ components. Such a manifold $M$ will have  $n$ disjoint torus boundaries. To compute the partition function, we must choose a basis for the first homology group of the boundary $\partial M$. A canonical choice of this basis is given by $\left\{(m_\alpha,\, \ell_\alpha) : \alpha=1,2,\ldots,n\right\}$ where $m_\alpha$ is the meridian and $\ell_\alpha$ is the longitude of the $\alpha^{\text{th}}$ torus. Next, we perform the ideal triangulation of the manifold $ M$. Suppose there are $j$ tetrahedra that are required for the triangulation. We will have $j$ parameters denoted by $z_1, z_2,\ldots,z_j$. The parameter $z_i$ of the $i^{\text{th}}$ tetrahedron 
is related to the shape parameter $Z_i$ as $Z_i=\exp(z_i)$ (see appendix \ref{triangulation} for details). In general $n \leq j$. The partition function will depend on the gluing matrices which will determine the way various tetrahedra are glued together. In addition, the partition function will also depend on the parameters $x_1, x_2, \ldots, x_n$, where $x_\alpha$ are the deformation parameters. These deformation parameters act to change the holonomy along the meridian of the complement manifold for the $\alpha^{\text{th}}$ torus \cite{gang}. For calculation purposes, it is convenient to collect these parameters into $\vec{x}$ and $\vec{z}$ as $j \times 1$ column matrices with entries as:
\begin{equation} 
\vec{x} = \{x_1, x_2,..., x_n, 0, 0,...,0\} \quad;\quad \vec{z} = \{z_1, z_2, z_3..., z_j\} ~.
\end{equation}
If we denote `ZPSL' to be the $\text{PSL}(2,\mathbb{C})$ Chern-Simons partition function, then we will have the following state-integral expression \cite{dimofte2013,dimofte20131,gang}:  
 \begin{equation}
\text{ZPSL}_{\hbar}(M\,;\, \vec{x})
= \frac{2}{\sqrt{\det B}}\int_{z_1} \int_{z_2} \cdots \int_{z_j} \frac{dz_1}{\sqrt{2 \pi \hbar}} \, \frac{dz_2}{\sqrt{2 \pi \hbar}} \, \ldots \frac{dz_j}{\sqrt{2 \pi \hbar}}\, \, \,     \exp(\frac{Q}{\hbar})\, \prod_{i=1}^j \Psi_{\hbar}(z_i)   ~.
\label{Dimofte integral}
 \end{equation}
In this expression, the Chern-Simons level $k$ has been analytically continued to $2\pi i/ \hbar$. In the integrand of \eqref{Dimofte integral}, the term $\Psi_{\hbar}(z)$ is the  simple partition function of an ideal tetrahedron which is defined as \cite{faddeev}: 
\begin{equation} \label{Q.D.L}
\Psi_{\hbar} (z) \equiv \prod_{r=1}^\infty \, \frac{1- \exp[\hbar r - z]} {1-\exp[-\frac{2\pi i}{\hbar} (2\pi i r - 2\pi i -z) ] } ~.
\end{equation}
The function $Q$ depends on various matrices and is given as:
\begin{align}
Q &= \frac{(\vec{z})^T B^{-1} A \vec{z}}{2}  + 2 (\vec{x})^T  D B^{-1} \vec{x}  - 2(\vec{z})^T B^{-1}\vec{x} +  (2\pi i +\hbar) (\vec{f})^T B^{-1} \vec{x}  \nonumber  \\
& + \frac{\left(2\pi i+\hbar\right)^2}{8} (\vec{f})^T B^{-1} \vec{\nu} - \frac{\left(2\pi i+\hbar\right)}{2}(\vec{z})^T B^{-1}  \vec{\nu}  \;.
\end{align}
Here $A, B, C, D, \vec{\nu}, \vec{\nu}'$ are the Neumann-Zagier matrices\footnote{The matrices $A$, $B$, $C$, $D$ are of order $j\times j$, where $j$ represents the number of tetrahedra in the triangulation of the manifold $M$. Additionally, $\vec{\nu}$ and $\vec{\nu}'$ are $j$-dimensional vectors, i.e. matrices of size $j\times 1$. The details of these matrices are given in the appendix and they can be computed numerically from the triangulation data generated using \texttt{SnapPy} \cite{SnapPy} for any given link complement.} The $\vec{f}$ comes from the combinatorial flattenings. The details of all these matrices are discussed in the appendix \ref{triangulation}. The integral in \eqref{Dimofte integral} is performed over the $z_i$ for all the tetrahedra. Note that given a link complement, the choice of an ideal triangulation and its gluing matrices is not unique. However, it was shown in \cite{dimofte20131}, that up to some ambiguity, the partition function \eqref{Dimofte integral} is independent of such choices. 

Up to this point, we have discussed how to compute the partition function \eqref{Dimofte integral} for the link-complement manifold $\mathbf S^3\backslash \mathcal L$. Next, we will illustrate the methodology to determine the partition function of the closed manifolds obtained by performing the Dehn filling on the toroidal boundaries of $\mathbf S^3\backslash \mathcal L$.

The Dehn filling can be performed on each of the torus boundaries of $\mathbf S^3\backslash \mathcal L$ separately. The Dehn filling on the $\alpha^{\text{th}}$ torus boundary is denoted by a pair of coprime integers $(p_\alpha, q_\alpha)$. It is achieved by filling the toroidal boundary with a solid torus along its two cycles in such a way that $p_\alpha m_\alpha + q_\alpha \ell_\alpha$ is contractible inside the solid torus.
Once the Dehn fillings on all the $n$ boundaries of $\mathbf S^3\backslash \mathcal L$ have been done, we obtain a closed 3-manifold:
\begin{equation}
    \hat{M} \equiv \left[\mathbf S^3 \backslash \mathcal L\right]_{(p_1, q_1),\ldots,(p_n, q_n)} ~.
\end{equation}
Since we know the partition function of $\mathbf S^3 \backslash \mathcal L$ from the state integral formula \eqref{Dimofte integral}, the partition function of the closed manifold $\hat{M}$ after applying the Dehn filling will be given as \cite{Bae}:
\begin{equation}
\text{ZPSL}_{\hbar}(\hat{M}) = \int_{x_1}\cdots \int_{x_n} \frac{dx_1}{\sqrt{2 \pi \hbar\, q_1}}   \ldots \frac{dx_n}{\sqrt{2 \pi \hbar\, q_n}}  \prod_{\alpha=1}^{n}\mathcal{K}_{p_{\alpha},\,q_{\alpha}}(x_\alpha)~ Z^{\text{PSL}(2,\mathbb{C})}_{\hbar}( M\,;\, \vec{x})~.
\label{Dehn_filling_formula}
\end{equation}
In the above expression $\mathcal{K}_{p_{\alpha},\,q_{\alpha}}(x_\alpha)$ provides the required Dehn twist on the solid torus which is being glued to $\alpha^{\text{th}}$ torus boundary. Its expression is given as:
\begin{align}
\mathcal{K}_{p_{\alpha},\,q_{\alpha}}(x_\alpha) &=
\exp\left[\frac{s_\alpha}{q_\alpha} \left(\frac{\pi ^2 }{\hbar}-\frac{\hbar }{4}\right) +\frac{p_\alpha\, x_\alpha^2}{\hbar\, q_\alpha} + \frac{2\pi i\, x_\alpha}{\hbar\, q_\alpha}\right] \sinh\left[\frac{x_\alpha-i \pi s_\alpha}{q_\alpha}\right] \nonumber \\
&- \exp\left[\frac{s_\alpha}{q_\alpha} \left(\frac{\pi ^2 }{\hbar}-\frac{\hbar }{4}\right) +\frac{p_\alpha\, x_\alpha^2}{\hbar\, q_\alpha} - \frac{2\pi i\, x_\alpha}{\hbar\, q_\alpha}\right] \sinh\left[\frac{x_\alpha+i \pi s_\alpha}{q_\alpha}\right]  ~,
\end{align}
where $r_\alpha$ and $s_\alpha$ are two integers such that 
\begin{equation}
\left(\begin{array}{cc}r_\alpha & s_\alpha \\ p_\alpha & q_\alpha  \end{array}\right) \in \text{PSL}(2,\mathbb{Z}) \quad \Longrightarrow \quad q_\alpha \,r_\alpha - p_\alpha \, s_\alpha = 1 ~.
\label{choicers}
\end{equation}
The equation \eqref{Dehn_filling_formula} evaluates the closed manifold partition function for a choice of Dehn filling of the link complement.  In the following subsection,  we will investigate the perturbative expansion of the partition function \eqref{Dehn_filling_formula}.
\subsection{Perturbative expansion of the state integral partition function}
Computing the full partition function from \eqref{Dehn_filling_formula} is intractable, even for the simplest of cases. Therefore, the partition function is obtained perturbatively in powers of $\hbar$, where we consider the parameter $\hbar$ as \cite{gang}:
\begin{equation} 
\hbar =2\pi i b^2  \in i \mathbb{R}_+ ~.
\end{equation} 
Using the perturbative expansions of $\Psi_{\hbar}(z_i)$ and $\mathcal{K}_{p_{\alpha},\,q_{\alpha}}(x_\alpha)$, the partition function is expanded as:
\begin{equation}
\text{ZPSL}_{\hbar}(\hat{M}) = \int \left[\frac{dx_\alpha}{\sqrt{2 \pi \hbar\, q_\alpha}} \right] \, \int \left[\frac{dz_i}{\sqrt{2 \pi \hbar}} \right]   \,\,\times \,\, \exp \left( \sum_{m=0}^\infty W_{m} (\vec{z}, \vec{x})\, \hbar^{m-1} \right) ~,
\label{expand}
\end{equation}
where the leading order term is \cite{gang}:
\begin{align}
    W_{0} &= \left(\frac{(\vec{z})^T B^{-1} A \vec{z}}{2}  + 2 (\vec{x})^T  D B^{-1} \vec{x}  - 2(\vec{z})^T B^{-1}\vec{x} +  2\pi i (\vec{f})^T B^{-1} \vec{x}  - \frac{\pi^2}{2} (\vec{f})^T B^{-1} \vec{\nu} - i\pi(\vec{z})^T B^{-1}  \vec{\nu} \right) \nonumber \\
    &+ \left(\sum_{i=1}^{k}\text{Li}_2(e^{-z_i})\right)+ \sum_{\alpha=1}^{n}\left(\frac{p_\alpha}{q_\alpha}x_\alpha^2+\frac{2\pi i}{q_\alpha}x_\alpha+\frac{\pi^2 s_\alpha}{q_\alpha}\right) ~.
    \label{W0expression}
\end{align}
In the above expression, the terms in the first line are coming from the expansion of $\exp(Q/\hbar)$. In the second line, the term in the first parentheses is due to the expansion of $\prod \Psi_{\hbar}(z_i)$. Finally, the term in the second parentheses is from the expansion of $\prod \mathcal{K}_{p_{\alpha},\,q_{\alpha}}(x_\alpha)$. The integrals over $z_i$ and $x_\alpha$ in the equation \eqref{expand} are performed using the saddle point approximation. The saddle point is obtained by solving the following set of $j+n$ equations:
\begin{align} 
\frac{\partial\, W_{0}(\vec{z}, \vec{x})}{\partial z_i} &= 0 \quad,\quad i=1,2,\ldots,j \nonumber \\ \frac{\partial\, W_{0}(\vec{z}, \vec{x})}{\partial x_\alpha} &= 0 \quad,\quad \alpha=1,2,\ldots,n ~.
\end{align}
Solving the above equations will give us $\vec{z} = \vec{z}_0$ and $\vec{x} = \vec{x}_0$. Thus the perturbative expansion of the partition function will be given as follows: 
\begin{equation} \label{}
\text{ZPSL}_{\hbar}(\hat{M}) = \exp[\frac1\hbar W_{0}(\vec{z}_0, \vec{x}_0)+W_{1}(\vec{z}_0, \vec{x}_0) +\hbar W_{2}(\vec{z}_0, \vec{x}_0)+\hbar^2 W_{3}(\vec{z}_0, \vec{x}_0)+\ldots ] ~.
\end{equation} 
The above partition function is traditionally written as:
\begin{equation} 
\text{ZPSL}_{\hbar}(\hat{M}) = \exp[\frac1\hbar S_{0} + S_{1} +\hbar\, S_{2}+\hbar^2 S_{3}+\ldots ] ~, \label{eqn:2.16}
\end{equation} 
where each term $S_m$ is a topological invariant of the manifold $\hat{M}$. Due to the non-unique choices of triangulation and gluing, these invariants are defined up to the following ambiguities \cite{gang}: 
\begin{equation}
S_{0} = S_{0} \, (\text{mod } \pi^2/6) \quad,\quad S_{1} = S_{1} \, (\text{mod } i\pi/4) \quad,\quad  S_{2} = S_{2} \, (\text{mod } 1/24) ~.  
\end{equation}
The leading order term $S_{0}$ gives the hyperbolic volume of the  closed manifold $\hat{M}$ as:
\begin{equation}
\text{Vol}(\hat{M}) = \text{Im}[S_{0}]  ~.  \label{asymphatM}
\end{equation}
Although we are focused only on the invariant $\text{Im}[S_{0}]$ in this work, the higher order invariants can be computed order by order as discussed in \cite{dimofte20131, gang}. We are now positioned to explore the topology of closed 3-manifolds derived from $\mathbf S^3  \backslash  (\mathcal K \# H)$ with suitable Dehn filling. In the following section, we will conduct a more detailed investigation of this aspect.
\section{\texorpdfstring{Dehn Filling of $\mathbf S^3 \backslash (\mathcal{K} \protect\# H)$}{Dehn Filling of S³\ K\#H}} \label{sec3}
\begin{figure}
\centering
\begin{minipage}{.5\textwidth}
  \centering
  \includegraphics[width=.65\linewidth]{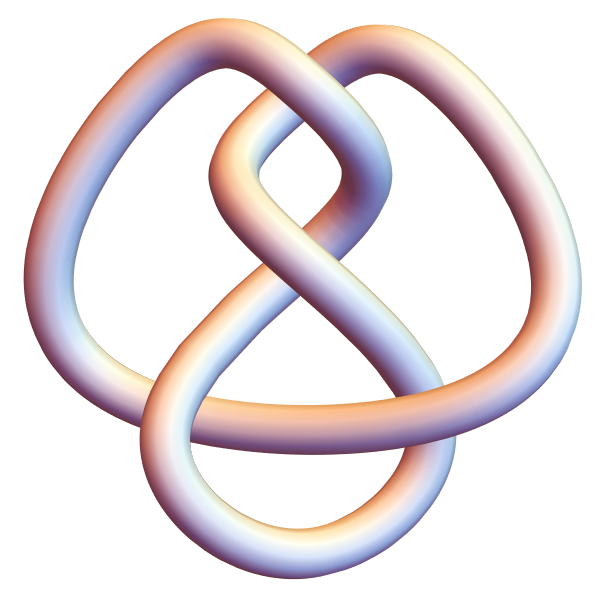}
  \captionof{figure}{Figure-eight knot ($4_1$ knot)}
  \label{fig8}
\end{minipage}%
\begin{minipage}{.5\textwidth}
  \centering
  \includegraphics[width=.9\linewidth]{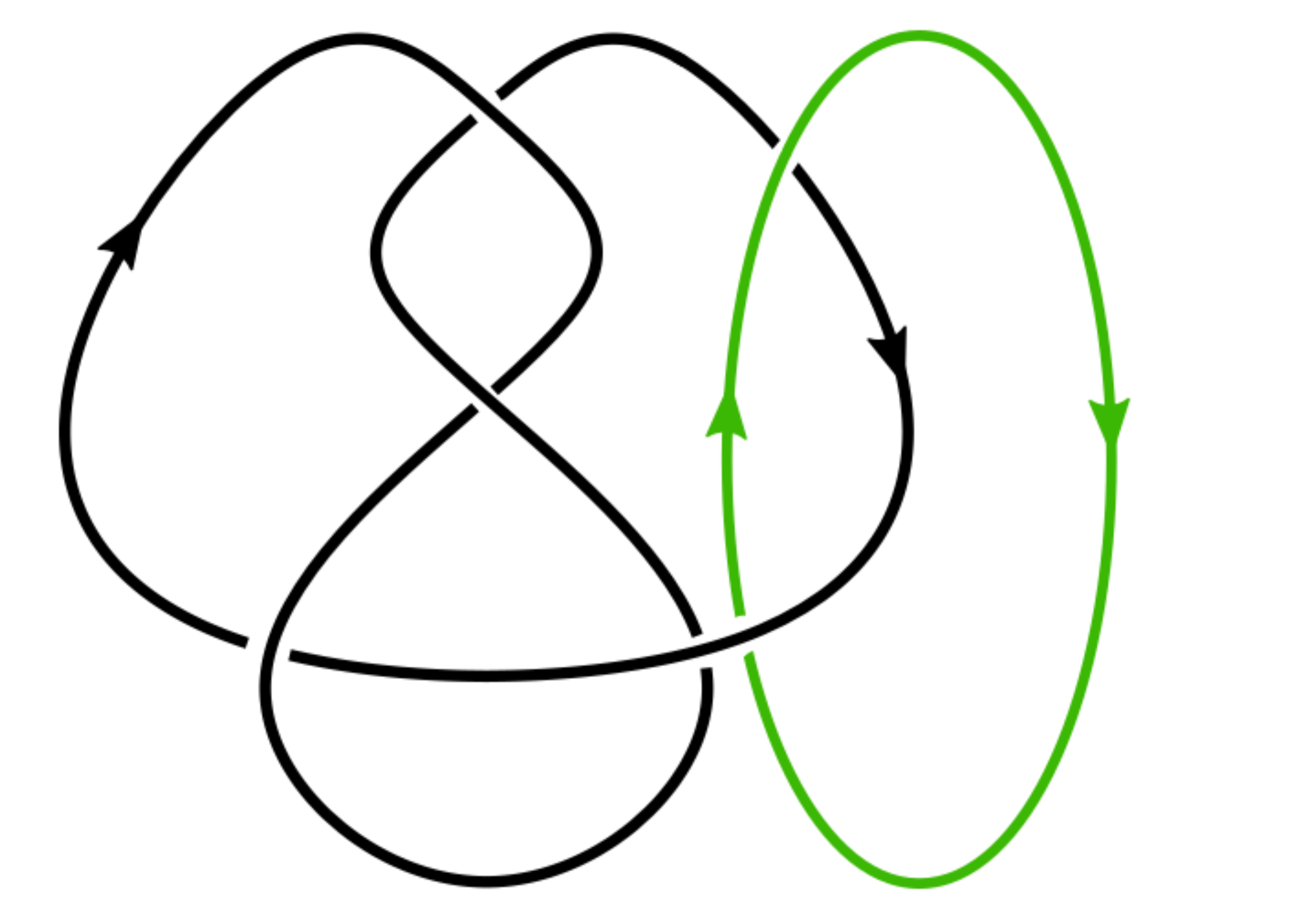}
  \captionof{figure}{$4_1 \# \text{Hopf link}$}
  \label{fig8+hopf}
\end{minipage}
\end{figure}

Here we consider closed 3-manifolds of type $\hat{M} = [\mathbf S^3 \backslash (\mathcal{K} \# H)]_{(p_1, q_1),(p_2, q_2)}$. This manifold is obtained by performing the Dehn fillings of the two toroidal boundaries of $\mathbf S^3 \backslash \mathcal (\mathcal {K} \# H)$. Here the link  $\mathcal K \# H$ is a two-component link, where one component is $\mathcal{K}$ and the other is unknot (a typical example is shown in the figure \ref{fig8+hopf}). Our convention is that we do the  $(p_1,q_1)$ Dehn filling on the boundary component associated with $\mathcal{K}$ and the $(p_2,q_2)$ Dehn filling on the boundary component associated with the unknot. We aim to compute the PSL(2,$\mathbb{C}$) Chern-Simons partition function of these manifolds and to compare its asymptotic behavior with that of the SO(3) Chern-Simons partition function of $M_{\mathcal{K}}$ as discussed in the section \ref{sec1}. We start with the simplest of the hyperbolic knot $\mathcal{K}=4_1$.
\subsection{\texorpdfstring{$\mathcal K=4_1$: The figure-eight knot}{}}\label{41}
The figure-eight or the $4_1$ knot is shown in the figure \ref{fig8}. Let us consider the manifold $\mathbf S^3 \backslash (4_1 \# H)$ where the link $4_1 \# H$ is shown in figure \ref{fig8+hopf}. We use the computer program \texttt{SnapPy} \cite{SnapPy} that helps in the study of topology and geometry of the 3-manifolds and can provide the triangulation data of the link complements. In the case of $\mathbf S^3 \backslash (4_1 \# H)$, six tetrahedra are required for the triangulation. As discussed earlier, the Neumman-Zagier matrices will be of order 6 and there will be $6+2 = 8$ integrals involved in the partition function. This leads to computational difficulty even for the low crossing knot like $4_1$. To circumvent this difficulty, 
we consider the manifold $[\mathbf S^3 \backslash (4_1 \# H)]_{(p_1, q_1)}$ which is a manifold obtained by the Dehn filling $(p_1,q_1)$ on the toroidal boundary associated with $4_1$ component. This manifold has a single torus boundary and we can give this manifold as an input in the \texttt{SnapPy}. Next choosing suitable Dehn fillings $(p_2,q_2)$ for the above manifold, we finally get the closed 3-manifold. The gluing equations and the Neumann-Zagier matrices for the manifold $[\mathbf S^3 \backslash (4_1 \# H)]_{(p_1, q_1)}$ can be read off from \texttt{SnapPy}. After that, the invariant $W_0$ and subsequently $S_0$ for the closed manifold $[\mathbf S^3 \backslash (4_1 \# H)]_{(p_1, q_1),(p_2,q_2)}$ can be obtained using \eqref{W0expression}.  

We will illustrate this procedure through an example. Note that we do not have a clear prescription or understanding of the choices of Dehn filling. After experimenting with \texttt{SnapPy}, we realized that the Dehn filling on the torus boundary of knot $\mathcal K$ component must be small. However, the second Dehn filling on the unknot component has to be large\footnote{It is evident from the Theorem $1.2\textit{A}$ of \cite{NEUMANN} that $\text{Vol}(\hat{M})=\text{Vol}(M)-\frac{1}{Q(p,q)}+O{(\frac{1}{p^4+q^4}})$, where $\hat{M}$ is the closed manifold obtained from $(p,q)$ Dehn filling of the torus boundary of $M$. Here $Q(p,q)$ is a positive function. Thus we can see that the volume of the closed manifold after the Dehn filling is less than the volume of the original manifold. When the Dehn filling is large, the $\text{Vol}(\hat{M})$ can be made equal to $\text{Vol}(M)$.  This theorem was used in \cite{gukov2005} to study the large Dehn filling of $\mathbf S^3 \backslash 4_1$.}  so that the leading order of the asymptotic expansion \eqref{eqn:2.16} matches with that of the hyperbolic volume (\ref{asymphatM}) of the $\mathbf S^3 \backslash \mathcal K$. Some of the results with various Dehn fillings are shown in the table \ref{DehnfillingsFig8}. 
\begin{table}[htb]
\centering
\begin{tabular}{|l||c|}
 \hline
 Dehn filling $(p_1,q_1),(p_2,q_2)$  & $\text{Im}[S_{0}] = \text{Volume of } [\mathbf S^3 \backslash (4_1 \# H)]_{(p_1, q_1),(p_2,q_2)}$  \\[0.1cm]
 \hline
$(1,1),(-1,1)$  & 0   \\
 $(1,1),(-1,2)$  & 0   \\
 $(1,1),(1,1)$  & 0 \\
 $(1,1),(1,2)$ & 0   \\
 $(1,1),(2,5)$  & $1.440699007$ \\
 $(1,1),(1500,7)$  &$2.02988321$  \\  \hline
\end{tabular}
\caption{The various Dehn fillings of the manifold $\mathbf S^3 \backslash (4_1 \# H)$. The first Dehn filling $(p_1,q_1)$ is performed on the $4_1$ component, and the second Dehn filling $(p_2,q_2)$ is performed on the unknot component. For each Dehn filling, the value of the \text{Im}[$S_0$], which equals the volume of the resulting closed manifold, is given in the second column.}
\label{DehnfillingsFig8}
\end{table}\\ 
\textbf{$\bullet$ Dehn filling $(1,1),(1500,7)$:} \\
As discussed earlier, we consider the manifold $[\mathbf S^3 \backslash (4_1 \# H)]_{(1, 1)}$ which is obtained by performing the $(1,1)$ Dehn filling on the $4_1$ component of $\mathbf S^3 \backslash (4_1 \# H)$. Next, using the \texttt{SnapPy}, we find that we need two tetrahedra to complete the triangulation of the manifold $[\mathbf S^3 \backslash (4_1 \# H)]_{(1, 1)}$. From the gluing equations generated by the \texttt{SnapPy}, one can read off the Neumann-Zagier matrices $A,B,C,D,\vec{\nu},\vec{\nu}'$. The steps for computing these matrices are presented in the appendix \ref{ex41}. The results are:
\begin{equation}
A=\left(
\begin{array}{cc}
	1 & 2 \\
	1 & 1 \\
\end{array}
\right), \,\,
B=\left(
 \begin{array}{cc}
 	0 & -3 \\
 	-1 & -1 \\
 \end{array}
 \right), \,\,
 C=\left(
\begin{array}{cc}
	1 & 0 \\
	0 & 2 \\
\end{array}
\right), \,\,
D=\left(
\begin{array}{cc}
	0 & -1 \\
	-2 & -2 \\
\end{array}
\right), \,\, \vec{\nu}= \left(
\begin{array}{c}
 0 \\
 0 \\
\end{array} \right), \,\,
\vec{\nu}'= \left(
\begin{array}{c}
 0 \\
 0 \\
\end{array} \right)
\end{equation}
We also find two combinatorial flattenings from where we can read off $\vec{f}$ as discussed in the appendix \ref{ex41} and it comes out to be: 
\begin{equation}
\vec{f}= \left(
\begin{array}{c}
 0 \\
 0 \\
\end{array} \right) ~.
\end{equation}
Using all this data, we can write the PSL(2,$\mathbb{C}$) Chern-Simons partition function of the closed 3-manifold $\hat{M}=[\mathbf S^3 \backslash (4_1 \# H)]_{(1, 1),(1500, 7)}$ in the integral form as:
\begin{equation}
\text{ZPSL}_\hbar(\hat{M}) = \frac{2}{\sqrt{\det B}} \int_{x}  \int_{z_1} \int_{z_2} \frac{dx}{\sqrt{14 \pi \hbar}} \frac{dz_1}{\sqrt{2 \pi \hbar}} \frac{dz_2}{\sqrt{2 \pi \hbar}}  \,     \exp(\frac{Q}{\hbar})\, \Psi_{\hbar}(z_1)\, \Psi_{\hbar}(z_2) \, \mathcal{K}_{1500,7}(x) ~.
\label{}
\end{equation}
We can chose $r=7,\,s=3$ following \eqref{choicers} and we get:
\begin{align}
\mathcal{K}_{1500,7}(x) &= e^{\frac{1500 x^2}{7 \hbar}+\frac{3 \pi ^2}{7 \hbar}-\frac{3 \hbar}{28}}
\left( e^{\frac{2\pi i\, x}{7\hbar}} \sinh\left[\frac{x-3i \pi}{7}\right] -e^{-\frac{2\pi i\, x}{7\hbar}} \sinh\left[\frac{x+3 i \pi}{7}\right] \right) \nonumber \\ 
Q(x,z_1,z_2) &= \frac{1}{3} \left(2 x^2-2 x z_1+2 x z_2-z_1^2-z_2^2-z_1 z_2\right) ~.
\end{align}
We can do the perturbative expansion of this partition function  as discussed earlier and we get the leading order term \eqref{W0expression} as:
\begin{equation}
W_0 = \text{Li}_2\left(e^{-z_1}\right)+\text{Li}_2\left(e^{-z_2}\right
   )+\frac{1}{21} \left(4514 x^2+14 x (z_2-z_1)+6 i \pi  x-7
   \left(z_1^2+z_1 z_2+z_2^2\right)+9 \pi ^2\right)    
\end{equation}
To get the saddle point, we need to find $(x^0,z_1^0,z_2^0)$ at which:
\begin{equation}
\frac{\partial\, W_{0}}{\partial z_1} = 0\, ,\, \frac{\partial\, W_{0}}{\partial z_2} = 0 \, ,\, \frac{\partial\, W_{0}}{\partial x} = 0 ~.
\end{equation}
The numerical value of the saddle point at which this happens is given below:
\begin{equation}
(x^0,z_1^0,z_2^0) =(2.58213\times10^{-6} - 0.00208541 i, 0.000832758 + 1.048 i, -0.000832348 + 1.04639 i) ~.
\end{equation}
The value of $W_0$ at this saddle point will give us the invariant $S_0$ as:
\begin{equation}
S_0 = W_{0}(x^0, z_1^0,z_2^0) = 5.875704487-2.02988321 i ~.
\end{equation}
The imaginary part of $S_0$ gives the volume of the closed manifold, and hence we get:
\begin{equation}
\text{Vol}(\hat{M}) = \text{Im}[S_{0}] = 2.029883213 = \text{Vol}(\mathbf S^3 \backslash 4_1) ~.  
\end{equation}
The volume of the manifold $\hat{M}$ can also be verified as discussed in the appendix \ref{appB}. Thus from \eqref{eqn:2.16}, we obtain the following asymptotic limit of the partition function:
\begin{equation} \label{41dehnasym}
\lim_{\hbar \to 0} \text{Im}\left[\hbar \times \ln \text{ZPSL}_\hbar(\hat{M})\right] = \text{Vol}(\mathbf S^3 \backslash 4_1) ~.
\end{equation}
When we compare this result with the result \eqref{cor1} obtained in \cite{ad}, we get an equivalence:
\begin{equation} \label{41asy}
\lim_{\hbar \to 0} \text{Im}\left[\hbar \times \ln \text{ZPSL}_\hbar(\hat{M})\right] = \lim_{k \to \infty} 2\pi \frac{\ln \text{ZSO}_k(M_{4_1})}{k} = \text{Vol}(\mathbf S^3 \backslash 4_1) ~.
\end{equation}
Since the parameters $\hbar$ and $k$ are related by the relation $h=2\pi i/k$, the above equation hints at the possibility that the manifold $\hat{M}$ can be topologically mapped to the manifold $M_{4_1}$. 
Recall that the manifold $[\mathbf S^3 \backslash (4_1 \# H)]_{(1, 1),(1500, 7)}$ is the closed 3-manifold obtained by performing the $(1,1),(1500,7)$ Dehn filling on $\mathbf S^3 \backslash (4_1 \# H)$. On the contrary, the manifold $M_{4_1}$ of \cite{ad} is the closed 3-manifold obtained by taking two copies of $\mathbf S^3 \backslash (4_1 \# H)$ and gluing them along the oppositely oriented boundaries. Therefore, the context of the origin of these manifolds is completely different, and yet our leading order analysis hints at the possibility of them being the same manifold (topologically).

\subsection{\texorpdfstring{$\mathcal K =5_2$ : The three-twist knot}{}}\label{52}
\begin{figure}
	\centering
		\includegraphics[width=0.32\textwidth]{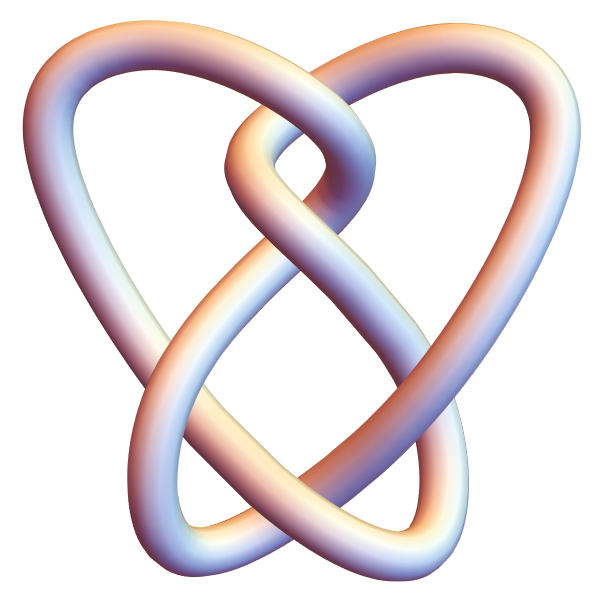}
	\caption{The three-twist knot ($5_2$ knot).}
	\label{52knot}
\end{figure}

The $5_2$ knot is shown in the figure \ref{52knot}. For the manifold $\mathbf S^3 \backslash (5_2 \# H)$ with two disjoint toroidal boundaries, we chose $(p_1,q_1)=(1,1)$ Dehn filling on the $5_2$ component torus boundary and tried various Dehn fillings on the unknot component. Some of the results are given in the table \ref{52data}.
\begin{table}[htb]
\centering
\begin{tabular}{|l||c|}
 \hline
 Dehn filling $(p_1,q_1),(p_2,q_2)$  & $\text{Im}[S_{0}] = \text{Volume of } [\mathbf S^3 \backslash (5_2 \# H)]_{(p_1, q_1),(p_2,q_2)}$  \\[0.1cm]
 \hline
 (1,1), (-1,1)   & 1.8435859723 \\
 (1,1), (-1,2)&   2.1030952907  \\
 (1,1), (-1,3)& 2.2726318636 \\
 (1,1), (2,1)& 2.4075734986\\
 (1,1), (3,1)& 2.6536786893   \\
 (1,1), (4,1)& 2.7330007496 \\
 (1,1), (291,7)& 2.8281220883\\
  (1,1), (1,1)& 0\\
   (1,1), (1,2)& 0\\
    (1,1), (1,6)& 0.9813688289\\
     (1,1), (1,7)&1.41406104417\\
     \hline
\end{tabular}
 \caption{The various Dehn fillings of the manifold $\mathbf S^3 \backslash (5_2 \# H)$. The first Dehn filling $(p_1,q_1)$ is performed on the $5_2$ component, and the second Dehn filling $(p_2,q_2)$ is performed on the unknot component. For each Dehn filling, the value of the \text{Im}[$S_0$], which equals the volume of the resulting closed manifold, is given in the second column.}
 \label{52data}
\end{table}
We see that the $(291,7)$ Dehn filling on the unknot component gives the volume that matches with that of the $\mathbf S^3\backslash 5_2$. This can also be verified as discussed in the appendix \ref{appB}. The explicit detail of the calculations for this Dehn filling is presented below.\\\\
\textbf{$\bullet$ Dehn filling $(1,1),(291,7)$ :} \\
Using \texttt{SnapPy}, we first perform the $(1,1)$ Dehn filling on the $5_2$ component of the manifold $\mathbf S^3 \backslash (5_2 \# H)$. This results in an intermediate manifold $\hat{M}=[\mathbf S^3 \backslash (5_2 \# H)]_{(1, 1)}$ with a single torus boundary. From \texttt{SnapPy}, we find that we need $3$ tetrahedra to complete the triangulation of this manifold. The \texttt{SnapPy} gives us the gluing equations, from which we can read off the Neumann-Zagier matrices as discussed in the appendix \ref{ex52} and we get:
\begin{equation*}
A=\left(
\begin{array}{ccc}
	2 & -1 & 0 \\
	0 & 1 & -1 \\
	-5 & 10 & -2 \\
\end{array}
\right), \,\,
B=\left(
\begin{array}{ccc}
	0 & 0 & -1 \\
	1 & 1 & 0 \\
	4 & 6 & 1 \\
\end{array}
\right), \,\,
 C=\left(
\begin{array}{ccc}
	1 & 0 & 1 \\
	1 & 0 & -5 \\
	1 & -2 & 1 \\
\end{array}
\right), \,\,
D=\left(
\begin{array}{ccc}
	1 & 1 & 0 \\
	3 & 1 & -2 \\
	-1 & -1 & 0 \\
\end{array}
\right) ~.
\end{equation*}
There are three combinatorial flattenings as discussed in the appendix \ref{ex52}. Thus we can read off all the single-column matrices as
\begin{equation}
\vec{f} = \left(
\begin{array}{c}
 0 \\
 0 \\
 -3 \\
\end{array}
\right), \quad \vec{\nu}= \left(
\begin{array}{c}
 0 \\
 1 \\
 -6 \\
\end{array}
\right), \quad
\vec{\nu}'= \left(
\begin{array}{c}
 -5 \\
 13 \\
 -1 \\
\end{array}
\right) ~.
\end{equation}
The calculations performed are similar to what was done in the previous case and we find that:
\begin{equation} \label{}
\lim_{\hbar \to 0} \text{Im}\left[\hbar \times \ln \text{ZPSL}_\hbar(\hat{M})\right] = \text{Vol}(\mathbf S^3 \backslash 5_2). 
\end{equation}
Comparing it with \eqref{cor1}, we get an equivalence:
\begin{equation}
\lim_{\hbar \to 0} \text{Im}\left[\hbar \times \ln \text{ZPSL}_\hbar(\hat{M})\right] = \lim_{k \to \infty} 2\pi \frac{\ln \text{ZSO}_k(M_{5_2})}{k} = \text{Vol}(\mathbf S^3 \backslash 5_2) ~.
\end{equation}
This information suggests the possibility that the 3-manifold $[\mathbf S^3 \backslash (5_2 \# H)]_{(1,1),(291,7)}$ might share the same topological properties as the manifold $M_{5_2}$, even though their origins and construction methods are quite different. The manifold $M_{5_2}$ is obtained by joining two copies of $\mathbf S^3 \backslash (5_2 \# H)$ along oppositely oriented boundaries while $[\mathbf S^3 \backslash (5_2 \# H)]_{(1,1),(291,7)}$ is obtained by the Dehn filling of $\mathbf S^3 \backslash (5_2 \# H)$.  This observation highlights an intriguing mathematical question about potential topological equivalence between these seemingly distinct 3-manifolds.

From table \ref{DehnfillingsFig8} and table \ref{52data}, we also observe some choice of Dehn fillings giving closed manifolds with vanishing $\text{Im}[S_{0}]$. These Dehn fillings are known as exceptional Dehn fillings and the corresponding manifolds are non-hyperbolic. We have also worked out similar computations for all the six-crossing knots which we present for completeness in the following subsection.
\subsection{Link complements and Dehn filling involving 6 crossing knots}
\begin{figure}
\centering
\begin{subfigure}{.33\textwidth}
  \centering
  \includegraphics[width=.85\linewidth]{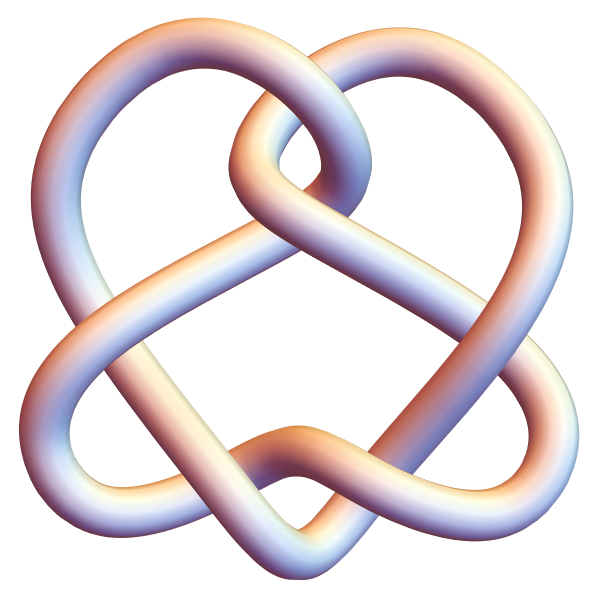}
  \caption{$6_1$ knot}
  \label{Knot61}
\end{subfigure}%
\begin{subfigure}{.33\textwidth}
  \centering
  \includegraphics[width=.85\linewidth]{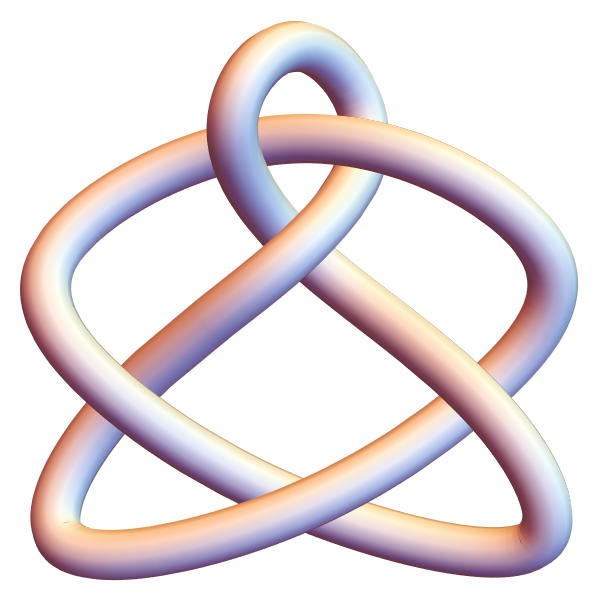}
  \caption{$6_2$ knot}
  \label{Knot62}
\end{subfigure}
\begin{subfigure}{.33\textwidth}
  \centering
  \includegraphics[width=.85\linewidth]{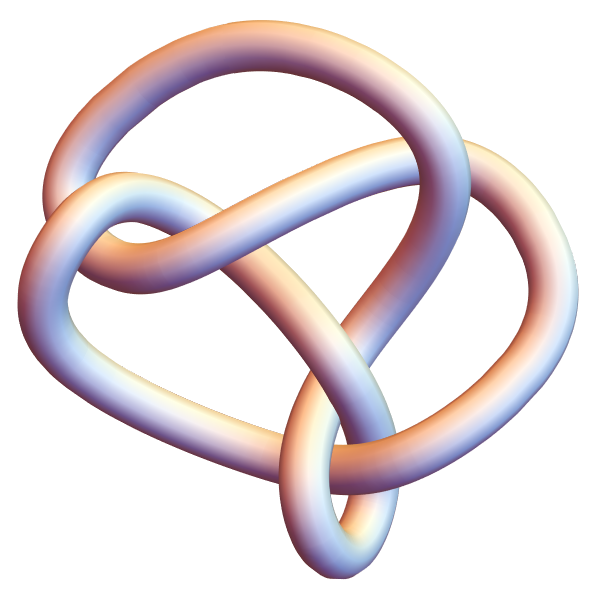}
  \caption{$6_3$ knot}
  \label{Knot63}
\end{subfigure}
\caption{Six crossing knots}
\label{6crossing}
\end{figure}

There are three knots with 6 crossings. These are named as $6_1$, $6_2$, and $6_3$ and are shown in the figure \ref{6crossing}. We will discuss each of these in the following. 
\subsubsection{\texorpdfstring{$\mathcal K=6_1$: The Stevedore knot}{}}\label{61}
As before, we tried various Dehn fillings on the manifold $\mathbf S^3 \backslash (6_1 \# H)$ to obtain the closed 3-manifold  $\hat{M} = [\mathbf S^3 \backslash (6_1 \# H)]_{(p_1, q_1),(p_2, q_2)}$. The volume of the resulting manifold for some of these Dehn fillings is given in the table \ref{61data}.
\begin{table}[htb]
\centering
\begin{tabular}{|l||c|}
 \hline
 Dehn filling $(p_1,q_1),(p_2,q_2)$  & $\text{Im}[S_{0}] = \text{Volume of } [\mathbf S^3 \backslash (6_1 \# H)]_{(p_1, q_1),(p_2,q_2)}$  \\[0.1cm]
 \hline
  (1,1), (-1,1)   & 0 \\
 (1,1), (-1,2)&    0  \\
 (1,1), (-1,3)& 0 \\
 (1,1), (-1,4)& 1.61046\\
 (1,1), (3,1)& 2.8112607    \\
 (1,1), (4,1)& 2.9604950 \\
 (1,1), (5,1)& 3.0317808\\
  (1,1), (1,1)& 0\\
   (1,1), (1,2)& 0\\
     (1,1), (1100,1)& 3.163960\\ 
     \hline
\end{tabular}
 \caption{The various Dehn fillings of the manifold $\mathbf S^3 \backslash (6_1 \# H)$. The first Dehn filling $(p_1,q_1)$ is performed on the $6_1$ component, and the second Dehn filling $(p_2,q_2)$ is performed on the unknot component. For each Dehn filling, the value of the \text{Im}[$S_0$], which equals the volume of the resulting closed manifold, is given in the second column.}
 \label{61data}
\end{table}
We see that the Dehn filling $(1,1), (1100,1)$ gives a hyperbolic volume that matches with the volume of $\mathbf S^3 \backslash 6_1$. Performing the $(1,1)$ Dehn filling on the $6_1$ toroidal component gives the manifold $[\mathbf S^3 \backslash (6_1 \# H)]_{(1, 1)}$ that has a  single torus boundary. According to \texttt{SnapPy}, four tetrahedra are required to complete its triangulation. Further, the \texttt{SnapPy} provides us with the gluing equations from which we can derive various gluing data. The gluing equations and the matrices $A,B,C,D,\vec{\nu},\vec{\nu}',\vec{f}$ have been explicitly obtained in the appendix \ref{ex61}. From the data obtained, we can state that for the manifold $\hat{M} = [\mathbf S^3 \backslash (6_1 \# H)]_{(1, 1),(1100, 1)}$, we have:
\begin{equation} \label{61sl2c}
\lim_{\hbar \to 0} \text{Im}\left[\hbar \times \ln \text{ZPSL}_\hbar(\hat{M})\right] = \text{Vol}(\mathbf S^3 \backslash 6_1) ~.
\end{equation}
Thus in view of \eqref{cor1}, we can write: 
\begin{equation} \label{61result}
\lim_{\hbar \to 0} \text{Im}\left[\hbar \times \ln \text{ZPSL}_\hbar(\hat{M})\right] = \lim_{k \to \infty} 2\pi \frac{\ln \text{ZSO}_k(M_{6_1})}{k} = \text{Vol}(\mathbf S^3 \backslash 6_1) ~.
\end{equation}
We have also verified the volume of the closed manifold $\hat{M}$ in the appendix \ref{appB}. Here again, we believe that there may be topological similarities between the closed 3-manifolds $[\mathbf S^3 \backslash (6_1 \# H)]_{(1, 1),(1100, 1)}$ and manifold $M_{6_1}$.
\subsubsection{\texorpdfstring{$\mathcal K=6_2$}{}}\label{62}
We tabulate the various Dehn fillings of the manifold $\mathbf S^3 \backslash (6_2 \# H)$ and the corresponding values of the volumes of the resultant manifold $\hat{M} = [\mathbf S^3 \backslash (6_2 \# H)]_{(p_1, q_1),(p_2, q_2)}$ in the table \ref{62data}.
\begin{table}[htb]
\centering
\begin{tabular}{|l||c|}
 \hline
 Dehn filling $(p_1,q_1),(p_2,q_2)$  & $\text{Im}[S_{0}] = \text{Volume of } [\mathbf S^3 \backslash (6_2 \# H)]_{(p_1, q_1),(p_2,q_2)}$  \\[0.1cm]
 \hline
(1,1), (1,9)   & 0 \\
 (1,1), (1,3)&   3.383197  \\
 (1,1), (1,6)& 2.50265931 \\
 (1,1), (3,1)& 4.343407    \\
 (1,1), (2791,1)& 4.4008325 \\
 (1,1), (5,1)& 3.16246\\
  (1,1), (1,1)& 3.77083\\
   (1,1), (1,2)& 3.60015\\ 
     \hline
\end{tabular}
 \caption{The various Dehn fillings of the manifold $\mathbf S^3 \backslash (6_2 \# H)$. The first Dehn filling $(p_1,q_1)$ is performed on the $6_2$ component, and the second Dehn filling $(p_2,q_2)$ is performed on the unknot component. For each Dehn filling, the value of the \text{Im}[$S_0$], which equals the volume of the resulting closed manifold, is given in the second column.}
 \label{62data}
\end{table}
We see that the Dehn filling  $(1,1), (2791,1)$  gives the hyperbolic volume that matches with that of $\mathbf S^3 \backslash 6_2$. We first perform the $(1,1)$ Dehn filling on the $6_2$ component leading to the intermediate manifold $[\mathbf S^3 \backslash (6_2 \# H)]_{(1,1)}$. \texttt{SnapPy} tells us that five tetrahedra are required to triangulate this manifold. The gluing equations along with the matrices $A,B,C,D,\vec{\nu},\vec{\nu}',\vec{f}$ are given in the appendix \ref{ex62}. 
Using this data, we can state that for the manifold $\hat{M} = [\mathbf S^3 \backslash (6_2 \# H)]_{(1, 1),(2791, 1)}$, we have:
\begin{equation}
\lim_{\hbar \to 0} \text{Im}\left[\hbar \times \ln \text{ZPSL}_\hbar(\hat{M})\right] = \text{Vol}(\mathbf S^3 \backslash 6_2) ~.
\end{equation}
Thus in view of \eqref{cor1}, we can write: 
\begin{equation}
\lim_{\hbar \to 0} \text{Im}\left[\hbar \times \ln \text{ZPSL}_\hbar(\hat{M})\right] = \lim_{k \to \infty} 2\pi \frac{\ln \text{ZSO}_k(M_{6_2})}{k} = \text{Vol}(\mathbf S^3 \backslash 6_2) ~.
\end{equation}
This hints at the possible topological equivalence of the closed 3-manifolds $[\mathbf S^3 \backslash (6_2 \# H)]_{(1, 1),(2791, 1)}$ and $M_{6_2}$. The volume of the former has also been verified numerically in the appendix \ref{appB}.
\subsubsection{\texorpdfstring{$\mathcal K=6_3$}{}}\label{63}
The results of various Dehn filling of the manifold $\mathbf S^3 \backslash (6_3 \# H)$ are tabulated in the table \ref{63data}.
\begin{table}[htb]
\centering
\begin{tabular}{|l||c|}
 \hline
 Dehn filling $(p_1,q_1),(p_2,q_2)$  & $\text{Im}[S_{0}] = \text{Volume of } [\mathbf S^3 \backslash (6_3 \# H)]_{(p_1, q_1),(p_2,q_2)}$  \\[0.1cm]
 \hline
 (1,1), (0,1)   & 0  \\
 (1,1), (1,1)&  4.059766  \\
 (1,1), (1,4)& 4.288608 \\
 (1,1), (573,17)& 5.69302\\
 (1,1), (1,3)&  4.163996    \\
 \hline
\end{tabular}
 \caption{The various Dehn fillings of the manifold $\mathbf S^3 \backslash (6_3 \# H)$. The first Dehn filling $(p_1,q_1)$ is performed on the $6_3$ component, and the second Dehn filling $(p_2,q_2)$ is performed on the unknot component. For each Dehn filling, the value of the \text{Im}[$S_0$], which equals the volume of the resulting closed manifold, is given in the second column.}
 \label{63data}
\end{table}
We observe that the closest Dehn filling such that the resulting closed manifold has the same hyperbolic volume as  that of $\mathbf S^3 \backslash 6_3$ is 
$(1,1), (573,17)$. The numerical calculation in \ref{appB} is also in numerical agreement with this.

We first perform the $(1,1)$ Dehn filling on $6_3$ component leading to an intermediate manifold $[\mathbf S^3 \backslash (6_3 \# H)]_{(1, 1)}$. This manifold requires six tetrahedra for its triangulation. The \texttt{SnapPy} results of the gluing equations and the combinatorial flattings are given in the appendix. Using these we can find out the matrices  $A,B,C,D,\vec{\nu},\vec{\nu}',\vec{f}$ as given in the appendix \ref{ex63}.
Thus for the manifold $\hat{M} = [\mathbf S^3 \backslash (6_3 \# H)]_{(1, 1),(573, 17)}$, we have:
\begin{equation}
\lim_{\hbar \to 0} \text{Im}\left[\hbar \times \ln \text{ZPSL}_\hbar(\hat{M})\right] = \text{Vol}(\mathbf S^3 \backslash 6_3) ~.
\end{equation}
Thus in view of \eqref{cor1}, we can write: 
\begin{equation}
\lim_{\hbar \to 0} \text{Im}\left[\hbar \times \ln \text{ZPSL}_\hbar(\hat{M})\right] = \lim_{k \to \infty} 2\pi \frac{\ln \text{ZSO}_k(M_{6_3})}{k} = \text{Vol}(\mathbf S^3 \backslash 6_3) ~.
\end{equation}
This hints at the topological equivalence of the closed 3-manifolds $[\mathbf S^3 \backslash (6_3 \# H)]_{(1, 1),(573, 17)}$ and $M_{6_3}$.
\section{Summary and Discussions}
\label{sec4}
Our study has focused on the $\text{PSL}(2, \mathbb{C})$ Chern-Simons partition function for closed 3-manifolds $[\mathbf S^3 \backslash (\mathcal{K} \# H)]_{(p_1, q_1),(p_2, q_2)}$ that are obtained by performing the Dehn fillings on the complement manifolds of knots connected to Hopf links, namely $\mathbf S^3 \backslash (\mathcal {K}\# H)$.

The 3-manifold $\mathbf S^3 \backslash (\mathcal {K}\# H)$ has two torus boundaries, one associated with the knot component and another with the unknot component involved in the connected sum. Our approach involves the ideal triangulation of $3$-manifold $\mathbf S^3 \backslash (\mathcal {K}\# H)$ and further Dehn filling on the torus boundary of the manifold. Dehn filling is a process of attaching solid tori to the torus boundaries along curves known as the longitude and meridian. For the manifold $[\mathbf S^3 \backslash (\mathcal{K} \# H)]_{(p_1, q_1),(p_2, q_2)}$, the $(p_1,q_1)$ Dehn filling is performed on the toroidal boundary associated with the $\mathcal{K}$ component and the $(p_2,q_2)$ Dehn filling is performed on the unknot component.

We have calculated the gluing equations using \texttt{SnapPy} and have derived the Neumann-Zagier matrices. Subsequently, we have computed the $\text{PSL}(2, \mathbb{C})$ partition function $\text{ZPSL}_\hbar$ for the closed $3$-manifolds $[\mathbf S^3 \backslash (\mathcal{K} \# H)]_{(p_1, q_1),(p_2, q_2)}$ using the state integral model \eqref{Dehn_filling_formula} and have computed the leading order term of the perturbative expansion of these partition functions. 

Our key result is that we were able to obtain the Dehn fillings for which the hyperbolic volume of the closed 3-manifold $[\mathbf S^3 \backslash (\mathcal{K} \# H)]_{(p_1, q_1),(p_2, q_2)}$ matches with that of $\mathbf S^3 \backslash \mathcal {K}$. In a parallel story, there exists a closed 3-manifold denoted as $M_\mathcal {K}$ (as discussed in \cite{ad}) which is obtained by gluing two copies of $\mathbf S^3 \backslash \mathcal (\mathcal {K} \# H)$ along their respective oppositely oriented boundaries. It was shown in \cite{ad} that the SO(3) partition functions of $M_\mathcal {K}$ show an asymptotic exponential behavior governed by the hyperbolic volume of $\mathbf S^3 \backslash \mathcal {K}$. When we consider the results of this work and the results of \cite{ad}, we get the following statement:\\\\
`The perturbative expansion ($\hbar \to 0$ limit) of the PSL(2,$\mathbb{C}$) Chern-Simons partition function of the closed 3-manifold $[\mathbf S^3 \backslash (\mathcal{K} \# H)]_{(p_1, q_1),(p_2, q_2)}$ matches with the perturbative expansion ($k \to \infty$ limit) of the SO(3) Chern-Simons partition function of the closed 3-manifold $M_{\mathcal{K}}$'. In particular, we were able to find the specific choice of the Dehn fillings $(p_1,q_1),(p_2,q_2)$ such that:
\begin{equation}
\lim_{\hbar \to 0} \text{Im}\left[\hbar \times \ln \text{ZPSL}_\hbar(M_{p_1,q_1}^{p_2,q_2})\right] = \lim_{k \to \infty} 2\pi \frac{\ln \text{ZSO}_k(M_{\mathcal{K}})}{k} = \text{Vol}(\mathbf S^3 \backslash \mathcal {K}) ~.
\end{equation}\\ 
We have verified this statement for all hyperbolic knots up to 6 crossings namely $\mathcal{K}=4_1, 5_2, 6_1, 6_2, 6_3$. 
Exploiting the connection between the $\text{PSL}(2,\mathbb{C})$ and SO(3) Chern-Simons theories and noting that the parameters $\hbar$ and $k$ are related by the relation $h=2\pi i/k$, the above result hints at the possibility that the manifold $[\mathbf S^3 \backslash (\mathcal{K} \# H)]_{(p_1, q_1),(p_2, q_2)}$, for the specific choice of the Dehn fillings, can be topologically mapped to the manifold $M_{\mathcal{K}}$. 

Beyond this quantitative result, we have no intuition or understanding of the equivalence of topology of $[\mathbf S^3 \backslash (\mathcal{K} \# H)]_{(p_1, q_1),(p_2, q_2)}$ and $M_{\mathcal {K}}$. Probably, computing the higher order terms $S_1,S_2$ (\ref{eqn:2.16}) in the perturbative expansions of the partition functions could be the direction to pursue in the future to confirm the equivalence of these manifolds. Further, many Dehn fillings could give the same hyperbolic volume \cite{gang}. It is not a priori clear whether they are topologically equivalent or distinct. The criteria for deciding the topological equivalence of manifolds is still an open problem.

From the quantum information point of view, the partition function $\text{ZSO}_k(M_{\mathcal{K}})$ is equal to the trace of the unnormalized reduced density matrix associated with the state $\ket{\mathcal{K}\# H}$. We believe that the topological equivalence of the manifold $M_\mathcal{K}$ with the Dehn-filled closed 3-manifolds corresponding to the link complements may give more insight into the structure of these density matrices. It will be interesting to pursue this direction and we leave this discussion for the future.
\\
\newline
{\textbf{Acknowledgments}} \\ \\
The authors would like to acknowledge Masahito Yamazaki and Mauricio Romo for useful discussion and suggestions. SD is supported by the ``SERB Start-Up Research Grant SRG/2023/001023''. The work of VKS is supported by ``Tamkeen under the NYU Abu
Dhabi Research Institute grant CG008 and ASPIRE Abu Dhabi under Project AARE20-
336''. BPM acknowledges the research grant for
faculty under the IoE Scheme (Number 6031) of Banaras Hindu University. PR would like to acknowledge the SPARC project grant ``SPARC/2019-2020/P2116/SL''. AD would like to thank Neil Hoffman for helpful discussions.

\appendix

\section{Triangulation of a manifold and the gluing datum}\label{triangulation}
\begin{figure}[htb]
    \centering
    \includegraphics[scale=.3]{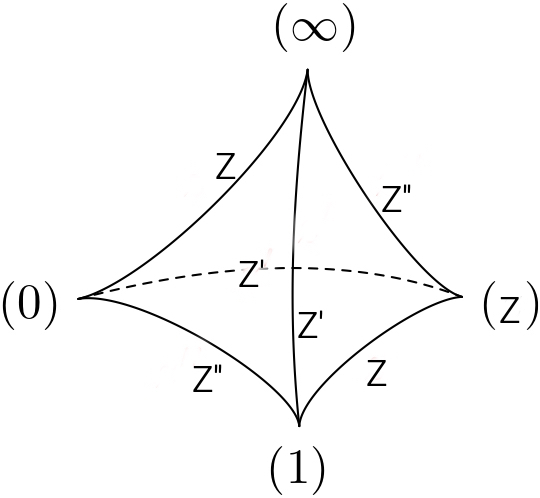}
    \caption{Ideal Tetrahedron}
    \label{figTetrahedron}
\end{figure}

We begin by reviewing the procedure \cite{dimofte2013} of ideal triangulation and the ways of computing the gluing rules involved in the ideal triangulation of a manifold $M$. Ideal triangulation involves decomposing the manifold $M$ into a collection of tetrahedra, where each tetrahedron is attached to others along their faces and edges, with all vertices removed. An oriented ideal tetrahedron $\Delta$ is characterized by $3$ complex shape parameters $(Z, Z', Z'')$, which are assigned to pairs of opposite edges as shown in the figure \ref{figTetrahedron}. 
These shape parameters satisfy
\begin{equation}
ZZ'Z'' = -1 \quad \text{and} \quad Z''+Z^{-1}-1=0
\label{tetexpP}
\end{equation}
This gives $Z'=1/(1-Z)$ and $Z''=1-Z^{-1}$. We define the ``quadrilateral type" of tetrahedron $\Delta$ by taking a cyclic permutation of the shape parameters $(Z, Z', Z'')$. It was shown in \cite{dimofte20131} that the equation  \eqref{tetexpP} remains invariant under the different choices of the quadrilateral type. Geometrically, both the $\mathrm{PSL}(2,\mathbb{C})$ structure and hyperbolic structure of the tetrahedron $\Delta$ can be determined by the shape parameters \cite{dimofte20131}. 
The shape parameters can also be written as:
\begin{equation} Z = \exp (\text{Torsion} +i \,\text{Dihedral angle})
\end{equation}
and similarly for  $Z'$, $Z''$.
We also define the logarithmic shape parameters $(z, z', z'')$ which are defined as 
$(Z, Z', Z'')=(e^z,e^{z'},e^{z''})$ and follow the condition that $z + z'+z''=i\pi$. This condition is because the sum of the dihedral angles going around any vertex is $2\pi$.
\subsection{Neumann-Zagier matrices and combinatorial flattenings}\label{gm}
The gluing rules of the ideal triangulation are specified by the matrices ${A, B, C, D, \vec{\nu}, \vec{\nu}'}$. These matrices are called Neumann-Zagier matrices and provide instructions for how the tetrahedra are combined, indicating how their faces and edges are glued together. We will discuss them briefly. Let $M$ be an oriented manifold with one torus boundary, and we do an ideal triangulation $ T=\{\Delta_1,\Delta_2,\ldots,\Delta_j\}$ of the manifold with a particular choice of quadrilateral type.
This choice of quadrilateral type and the orientation of manifold $M$, allows us to specify the opposite edges of each tetrahedron $\Delta_i$ with the variables $(z_i,z_i',z_i'')$.
Euler characteristic shows that if the triangulation involves $j$ tetrahedra, then the triangulation has 
$j$ edges $E_1, E_2,\ldots, E_j$. Suppose $J_{Ii}$ (respectively $J_{Ii}'$, $J_{Ii}''$) represent the number of times an edge of tetrahedron $\Delta_i$ with parameter $z_i$ (respectively, $z_i'$, $z_i''$) goes around the edge $E_I$ in the triangulation. Then the gluing equations for various edges will be:
\begin{equation} 
\sum_{i=1}^j \left(J_{Ii}z_i+J_{Ii}'z_i'+J_{Ii}''z_i''\right) = 2\pi i\, ,\quad \text{where} \quad I=1,2,\ldots,j ~. \label{glue} \end{equation}
Here $J_{Ii}$, $J_{Ii}'$ and $J_{Ii}''$ can take values 0, 1 or 2. The gluing equations are not all independent because of the relations \eqref{tetexpP} and we can reduce the number of shape parameters in the gluing equations. 

An oriented simple closed curve on the boundary of $M$ also gives rise to a gluing equation. So if $\mu$ and $\ell$ are the meridian and longitudes of the toroidal boundary of $M$ respectively, then the gluing equation associated with them can be written as \cite{dimofte20131}:
\begin{align} 
\sum_{i=1}^j \left(J_{j+1,\,i} \,z_i+J_{j+1,\,i}'\, z_i' + J_{j+1,\,i}''\,z_i'' \right) &= 0 \quad \text{(for meridian)} \label{merid} \\  
\sum_{i=1}^j \left(J_{j+2,\,i} \,z_i+J_{j+2,\,i}'\, z_i' + J_{j+2,\,i}''\,z_i'' \right) &= 0 \quad \text{(for longitude)}~.
\label{longi} 
\end{align}
Here $J_{j+1,\,i}\,\,,\,J_{j+1,\,i}'\,\,,\,J_{j+1,\,i}''\,\,,\, J_{j+2,\,i}\,\,,\,J_{j+2,\,i}'\,\,,\,J_{j+2,\,i}''\in \mathbb Z$. 

All the independent gluing equations coming from \eqref{glue} along with the meridian equation \eqref{merid} can be conveniently given as (after removing shape parameter $z_i'$):
\begin{equation} 
	\sum_{i=1}^j \left( A_{b i}\,z_i+ B_{b i}\,z_i''\right) = i \pi\, \nu_b
	\quad,\quad \text{where} \,\, b=1,2,\ldots,j ~. 
	\label{NZlin} 
\end{equation}
The above equation can be written as a matrix equation $ A\vec{z}+ B\vec{z}''=i\pi \vec{\nu}$ where $A,B$ are $j\times j$ matrices and $\vec{z},\vec{z}'',\vec{\nu}$ are $j\times 1$ column matrices.
To remove the edge equations, we must ensure that the condition $\text{det}(B) \neq 0$ is always satisfied. Apart from these matrices, there are three more matrices $C,D, \vec{\nu}'$ that encode the longitude equation \eqref{longi} and can be written as $C\vec{z}+ D\vec{z}''=i\pi \vec{\nu}'$. There is an additional constraint that the matrices $A,B,C,D$ form a symplectic block matrix i.e. \cite{NEUMANN, mathews2022symplectic}
\begin{equation}
    \begin{bmatrix}
  A & B \\
  C & D
\end{bmatrix} = \text{symplectic matrix} ~.
\label{ABCDSymplectic}
\end{equation}
The matrices $A, B, C, D,\vec{\nu},\vec{\nu}'$ are called the Neumann-Zagier matrices. This set of matrices can be obtained from the triangulation data produced by \texttt{SnapPy} \cite{SnapPy}.
 
When the manifold $M$ has a boundary, we must also consider deformations of the boundary holonomy. Thus the gluing equations in such a case will also be modified. If $M$ has $n$ number of toroidal boundaries, we introduce the deformation parameters $x_1, x_2, \ldots, x_{n}$ for each boundary. The modified gluing equations will be (more details can be found in \cite{gang}):
\begin{align}
\sum_{i=1}^j A_{bi} \, z_i+ B_{bi} \, z''_i=2x_b+i\pi \nu_b \quad,\quad \text{where} \,\, b=1,2,\ldots,j ~.
\label{gluingx1}
\end{align}

For an ideal triangulation $T$ of $M$, we define combinatorial flattening \cite{Neumann1992}
as a set of $3j$ integers $F_i=(f_i,f_i',f_i'') \in \mathbb{Z}^3$ for $i=1,\dots,j$
that satisfy:
\begin{align} 
        f_i + f_i' + f_i'' &= 1, \qquad i = 1,...,j \nonumber \\
        \sum_{i=1}^j \big(J_{Ii}f_i + J_{Ii}'f_i' + J_{Ii}''f_i''\big) &= \begin{cases} 
            2 & \text{for } I = 1,...,j, \\
            0 & \text{for } I = j+1 
        \end{cases} \label{flatG}
    \end{align} 
We can eliminate all $f_i'$ from the above equations so that the combinatorial flattenings can be expressed in terms of $f_i$ and $f_i''$ for $i=1,2,\ldots,j$. The integers $f_i$ can be collected into a matrix $\vec{f} = \left(
\begin{array}{cccc}
 f_1 & f_2 & \ldots & f_j
\end{array}
\right)^T$. Similarly, the integers $f_i''$ can be collected into a matrix $\vec{f}''=\left(
\begin{array}{cccc}
 f_1'' & f_2'' & \ldots & f_j''
\end{array}
\right)^T$. These matrices satisfy:
\begin{equation} 
	  A\vec{f}+B \vec{f}'' = \vec{\nu} \quad,\quad  C \vec{f}+D \vec{f}'' = \vec{\nu}' ~. 
\label{flatAB}
\end{equation}
\subsection{Some examples} \label{Example}
Here we present some examples of how to use the gluing equations generated by \texttt{SnapPy} to find various gluing data discussed in the previous subsections.
\subsubsection{\texorpdfstring{Triangulation of $[\mathbf S^3 \backslash (4_1 \# H)]_{(1,1)}$}{}} \label{ex41}
The manifold $[\mathbf S^3 \backslash (4_1 \# H)]_{(1,1)}$ is obtained by performing the $(1,1)$ Dehn filling on the toroidal boundary associated with the $4_1$ component of the manifold $\mathbf S^3 \backslash (4_1 \# H)$. From the \texttt{SnapPy}, we find that two tetrahedra are required for its ideal triangulation, so $T=\{\Delta_1, \Delta_2\}$. There are two edges along with meridian and longitude. The gluing equations can be read off from \texttt{SnapPy} and are given as:
\begin{align}
   z'_1+2 z''_1+ z'_2 + 2 z''_2 &= 2 i \pi \qquad \text{(edge 1)}  \nonumber \\
   2 z_1 + z'_1+2 z_2+z'_z &= 2 i \pi  \qquad \text{(edge 2)} \nonumber \\
   2 z_1+z'_1+z''_1+z_2-z'_2-4 z''_2 &= 0  \qquad \text{(meridian)} \nonumber  \\
   z''_2-z_1 &= 0 \qquad \text{(longitude)} 
   \label{gluingeq41}
\end{align} 
The above set of equations is not independent due to the relation between the shape parameters: $z_1+z'_1+z''_1=i \pi$ and $z_2+z'_2+z''_2=i \pi$. The first three equations of \eqref{gluingeq41} can be reduced to the following equations:
\begin{equation}
z_1 + 2 z_2 - 3 z''_2 = 0 \quad;\quad z_1 + z_2 - z''_1 - z''_2 = 0 ~.  
\end{equation}
This pair of equations can be written in a matrix form as $A\vec{z}+ B\vec{z}''=i\pi \vec{\nu}$ from where we can read off the matrices as:
\begin{equation}
    A=\left(
\begin{array}{cc}
	1 & 2 \\
	1 & 1 \\
\end{array}
\right)\ , \ B=\left(
 \begin{array}{cc}
 	0 & -3 \\
 	-1 & -1 \\
 \end{array}
 \right)\ , \ 
\vec{\nu}= \left(
\begin{array}{c}
 0 \\
 0 \\
\end{array} \right) ~.
\end{equation}
Next, we use the longitude equation of \eqref{gluingeq41} which reads as $z_1 - z''_2 = 0$. This equation along with the constraint \eqref{ABCDSymplectic} can give the matrices $C,D,\vec{\nu}'$ satisfying $C\vec{z}+ D\vec{z}''=i\pi \vec{\nu}'$ and can be written as:
\begin{equation*}
    C=\left(
\begin{array}{cc}
	1 & 0 \\
	0 & 2 \\
\end{array}
\right)\ ,\ D=\left(
\begin{array}{cc}
	0 & -1 \\
	-2 & -2 \\
\end{array}
\right)\ , \ \vec{\nu}'=\left(
\begin{array}{c}
 0 \\
 0 \\
\end{array}
\right)~  .
\end{equation*}
Combinatorial flattenings can be obtained using the relations \eqref{flatG} and \eqref{flatAB} and come out to be: 
\begin{equation}
F_1=(f_1,f_1',f_1'')=(0,1,0) \quad;\quad F_2=(f_2,f_2',f_2'')=(0,1,0) ~.
\end{equation}
This fixes the following matrices:
\begin{equation}
\vec{f}=\left(
\begin{array}{c}
 0 \\
 0 \\
\end{array}
\right);
\qquad
\vec{f}''=\left(
\begin{array}{c}
 0 \\
 0 \\
\end{array}
\right)~.
\end{equation}
\subsubsection{\texorpdfstring{Triangulation of $[\mathbf S^3 \backslash (5_2 \# H)]_{(1,1)}$}{}} \label{ex52}
The manifold $[\mathbf S^3 \backslash (5_2 \# H)]_{(1,1)}$ is obtained by performing the $(1,1)$ Dehn filling on the toroidal boundary associated with the $5_2$ component of the manifold $\mathbf S^3 \backslash (5_2 \# H)$. From the \texttt{SnapPy}, we find that three tetrahedra are required for its ideal triangulation, so $T=\{\Delta_1, \Delta_2,\Delta_3\}$. There are three edges along with the meridian and longitude. The gluing equations can be read off from \texttt{SnapPy} and are given as:
\begin{align}
  2 z_1' + z_2' + z_2'' + z_3 + z_3' &= 2 \pi i \qquad \text{(edge 1)}  \nonumber \\
   z_1'' + z_2 + z_2'' + z_3' + z_3'' &= 2 \pi i \qquad \text{(edge 2)}  \nonumber \\
  2 z_1' + z_1'' + z_2 + z_2' + z_3 + z_3''  &= 2 \pi i \qquad \text{(edge 3)}  \nonumber \\
  -4 z_1 + z_1' + 5 z_1'' + 5 z_2 - 5 z_2' + z_2'' - 4 z_3 - 2 z_3' - z_3''  &= 0 \qquad \text{(meridian)}  \nonumber \\
  z_1 - z_1'' - z_2 + z_2' + z_3 &= 0 \qquad \text{(longitude)} ~.
   \label{gluingeq52}
\end{align}
The above set of equations is not independent due to the relation between the shape parameters: $z_a+z'_a+z''_a=i \pi$ for $a=1,2,3$. The first four equations of \eqref{gluingeq52} can be reduced to the following equations:
\begin{equation}
2 z_1 - z_2 - z_3''=0 \,; \quad  z_1'' + z_2 + z_2'' - z_3=i \pi \,; \quad   4 z_1'' + 10 z_2 + 6 z_2'' + z_3'' - 5 z_1 + 2 z_3= 6 \pi i ~. 
\end{equation}
This set of equations can be written in a matrix form as $A\vec{z}+ B\vec{z}''=i\pi \vec{\nu}$ from where we can read off the matrices as:
\begin{equation}
    A=\left(
\begin{array}{ccc}
	2 & -1 & 0 \\
	0 & 1 & -1 \\
	-5 & 10 & -2 \\
\end{array}
\right)\ , \ B=\left(
\begin{array}{ccc}
	0 & 0 & -1 \\
	1 & 1 & 0 \\
	4 & 6 & 1 \\
\end{array}
\right)\ , \ 
\vec{\nu}= \left(
\begin{array}{c}
 0 \\
 1 \\
 6 \\
\end{array}
\right) ~.
\end{equation}
Next, we use the longitude equation of \eqref{gluingeq52} which reads as $  z_1 + z_3 - z_1'' - 2 z_2 - z_2''=-i \pi$. This equation along with the constraint \eqref{ABCDSymplectic} can give the matrices $C,D,\vec{\nu}'$ satisfying $C\vec{z}+ D\vec{z}''=i\pi \vec{\nu}'$ and can be written as:
\begin{equation*}
    C=\left(
\begin{array}{ccc}
	1 & 0 & 1 \\
	1 & 0 & -5 \\
	1 & -2 & 1 \\
\end{array}
\right)\ ,\ D=\left(
\begin{array}{ccc}
	1 & 1 & 0 \\
	3 & 1 & -2 \\
	-1 & -1 & 0 \\
\end{array}
\right)\ , \ \vec{\nu}'=\left(
\begin{array}{c}
 -5 \\
 13 \\
 -1 \\
\end{array}
\right) ~  .
\end{equation*}
Combinatorial flattenings can be obtained using the relations \eqref{flatG} and \eqref{flatAB} and come out to be: 
\begin{equation}
F_1=(0,1,0) \ , \ F_2=(0,0,3) \ , \ F_3=(-3,4,0) ~.
\end{equation}
This fixes the following matrices:
\begin{equation}
\vec{f}=\left(
\begin{array}{c}
 0 \\
 0 \\
 -3 \\
\end{array}\right); \qquad  \vec{f}''=
\left(
\begin{array}{c}
 0 \\
 3 \\
 0 \\
\end{array}
\right) ~.
\end{equation}
\subsubsection{\texorpdfstring{Triangulation of $[\mathbf S^3 \backslash (6_1 \# H)]_{(1,1)}$}{}} \label{ex61}
The manifold $[\mathbf S^3 \backslash (6_1 \# H)]_{(1,1)}$ is obtained by performing the $(1,1)$ Dehn filling on the toroidal boundary associated with the $6_1$ component of the manifold $\mathbf S^3 \backslash (6_1 \# H)$. From the \texttt{SnapPy}, we find that four tetrahedra are required for its ideal triangulation, so $T=\{\Delta_1, \Delta_2,\Delta_3,\Delta_4\}$. There are four edges along with the meridian and longitude. The gluing equations can be read off from \texttt{SnapPy} and are given as:
\begin{align}
   z_1' + z_1'' + z_2 + z_2' + 2 z_3 + z_4'' &= 2 \pi i \qquad \text{(edge 1)}  \nonumber \\
   Z1 + Z22 + Z333 + Z4 &= 2 \pi i \qquad \text{(edge 2)}  \nonumber \\
   z_1 + z_1'' + z_2'' + z_4' + z_4'' &= 2 \pi i \qquad \text{(edge 3)}  \nonumber \\
   z_1' + z_2 + z_2'' + 2 z_3' + z_3'' + z_4 + z_4' &= 2 \pi i \qquad \text{(edge 4)}  \nonumber \\
  z_1' + z_2 + z_2' + z_2'' + z_3' - z_3'' - 3 z_4 &= 0 \qquad \text{(meridian)}  \nonumber \\
  -z_2'' + z_3 &= 0 \qquad \text{(longitude)} ~.
   \label{gluingeq61}
\end{align}
The above set of equations is not independent due to the relation between the shape parameters: $z_a+z'_a+z''_a=i \pi$ for $a=1,2,3,4$. The first five equations of \eqref{gluingeq61} can be reduced to the following equations:
\begin{equation}
z_1 + z_2'' - 2 z_3 -z_4''=0 , \quad  z_1 + z_3'' + z_4 - z_2 - z_2''=i \pi, \quad  z_1 + z_1'' + z_2''- z_4=i \pi, \quad z_1 + z_1'' + z_3 + 2 z_3'' + 3 z_4 = 3 \pi i  
\end{equation}
This set of equations can be written in a matrix form as $A\vec{z}+ B\vec{z}''=i\pi \vec{\nu}$ from where we can read off the matrices as:
\begin{equation}
    A=\left(
	\begin{array}{cccc}
		1 & 0 & -2 & 0 \\
		1 & -1 & 0 & 1 \\
		1 & 0 & 0 & -1 \\
		1 & 0 & 1 & 3 \\
	\end{array}
	\right)\ , \ B=\left(
	\begin{array}{cccc}
		0 & 1 & 0 & -1 \\
		0 & -1 & 1 & 0 \\
		1 & 1 & 0 & 0 \\
		1 & 0 & 2 & 0 \\
	\end{array}
	\right)\ , \ 
\vec{\nu}= \left(
\begin{array}{c}
 0 \\
 1 \\
 1 \\
 3\\
\end{array}
\right) ~.
\end{equation}
Next, we use the longitude equation of \eqref{gluingeq61} which reads as $z_2'' - z_3=0$. This equation along with the constraint \eqref{ABCDSymplectic} can give the matrices $C,D,\vec{\nu}'$ satisfying $C\vec{z}+ D\vec{z}''=i\pi \vec{\nu}'$ and can be written as:
\begin{equation*}
    C=\left(
	\begin{array}{cccc}
		0 & 0 & 0 & 2 \\
		0 & 0 & 0 & 0 \\
		0 & 0 & 1 & -2 \\
		0 & 0 & -1 & 0 \\
	\end{array}
	\right)\ ,\ D=\left(
	\begin{array}{cccc}
		0 & 0 & 0 & 0 \\
		0 & -2 & 0 & 0 \\
		2 & 1 & 0 & 0 \\
		0 & 1 & 0 & 0 \\
	\end{array}
	\right)\ , \ \vec{\nu}'=\left(
\begin{array}{c}
 0 \\
 0 \\
 2 \\
 0\\
\end{array}
\right)  ~  .
\end{equation*}
Combinatorial flattenings can be obtained using the relations \eqref{flatG} and \eqref{flatAB} and come out to be: 
\begin{equation*}
    F_1 = \left(0, 0, 1\right),\,\
F_2 = \left(0, 1, 0\right),\,\
F_3 = \left(0, 0, 1\right),\,\
F_4 = \left(0, 1, 0\right) ~.
\end{equation*}
This fixes the following matrices:
\begin{equation}
\vec{f}=\left(
\begin{array}{c}
 0 \\
 0 \\
 0 \\
 0
\end{array}
\right); \qquad
\vec{f}''=\left(
\begin{array}{c}
 1 \\
 0 \\
 1 \\
 0
\end{array}\right)~.
\end{equation}
\subsubsection{\texorpdfstring{Triangulation of $[\mathbf S^3 \backslash (6_2 \# H)]_{(1,1)}$}{}} \label{ex62}
The manifold $[\mathbf S^3 \backslash (6_2 \# H)]_{(1,1)}$ is obtained by performing the $(1,1)$ Dehn filling on the toroidal boundary associated with the $6_2$ component of the manifold $\mathbf S^3 \backslash (6_2 \# H)$. From the \texttt{SnapPy}, we find that five tetrahedra are required for its ideal triangulation, so $T=\{\Delta_1, \Delta_2,\Delta_3,\Delta_4,\Delta_5\}$. There are five edges along with the meridian and longitude. The gluing equations can be read off from \texttt{SnapPy} and are given as:
\begin{align}
   z_1 + z_3' + z_4 + z_5' &= 2 \pi i \qquad \text{(edge 1)}  \nonumber \\
   z_1 + z_2' + z_2'' + z_3' + z_3'' + z_4'' &= 2 \pi i \qquad \text{(edge 2)}  \nonumber \\
   z_1' + z_2 + z_2' + z_3 + z_5 + z_5'' &= 2 \pi i \qquad \text{(edge 3)}  \nonumber \\
   z_1 + z_1'' + z_2 + z_3'' + z_4' + z_4'' + z_5 &= 2 \pi i \qquad \text{(edge 4)}  \nonumber \\
   z_1'' + z_2'' + z_3 + z_4 + z_4' + z_5' + z_5'' &= 2 \pi i \qquad \text{(edge 5)}  \nonumber \\
  -2 z_1' + 7 z_2 + z_2' + z_2'' - 9 z_3 + 8 z_3'' - 7 z_4 - z_4' + 8 z_5 - 
 6 z_5'' &= 0 \qquad \text{(meridian)}  \nonumber \\
  -z_2 + z_3 - z_3'' + z_4 - z_5 + z_5'' &= 0 \qquad \text{(longitude)} ~.
   \label{gluingeq62}
\end{align}
The above set of equations is not independent due to the relation between the shape parameters: $z_a+z'_a+z''_a=i \pi$ for $a=1,2,3,4,5$. The first six equations of \eqref{gluingeq62} can be reduced to the following equations:
\begin{align}
   z_1 + z_4 - z_3 -z_3'' - z_5 - z_5''& =0    \nonumber \\
   z_1 + z_4'' - z_2 + z_3 & =0  \nonumber \\
   z_1 + z_1'' + z_2'' - z_3 - z_5 - z_5''&=0   \nonumber \\
   z_1 + z_4 - z_2 - z_3'' - z_5&=0   \nonumber \\
      2 z_1 + 2 z_1'' + 6 z_2 + 8 z_3'' + z_4'' + 8 z_5 - 
 9 z_3 - 6 z_4 -6 z_5''&=2 \pi i   ~.
\end{align}
This set of equations can be written in a matrix form as $A\vec{z}+ B\vec{z}''=i\pi \vec{\nu}$ from where we can read off the matrices as:
\begin{equation}
    A=\left(
	\begin{array}{ccccc}
		1 & 0 & -1 & 1 & -1 \\
		1 & -1 & -1 & 0 & 0 \\
		1 & 0 & -1 & 0 & -1 \\
		1 & -1 & 0 & 1 & -1 \\
		2 & 6 & -9 & -6 & 8 \\
	\end{array}
	\right)\ , \ B=\left(
	\begin{array}{ccccc}
		0 & 0 & -1 & 0 & -1 \\
		0 & 0 & 0 & 1 & 0 \\
		1 & 1 & 0 & 0 & -1 \\
		0 & 0 & -1 & 0 & 0 \\
		2 & 0 & 8 & 1 & -6 \\
	\end{array}
	\right)\ , \ 
\vec{\nu}= \left(
\begin{array}{c}
 0 \\
 0 \\
 0 \\
 0\\
 2\\
\end{array}
\right) ~.
\end{equation}
Next, we use the longitude equation of \eqref{gluingeq62} which reads as $z_2 + z_3'' + z_5 - z_3 - z_4 - z_5''=0$. This equation along with the constraint \eqref{ABCDSymplectic} can give the matrices $C,D,\vec{\nu}'$ satisfying $C\vec{z}+ D\vec{z}''=i\pi \vec{\nu}'$ and can be written as:
\begin{equation*}
    C=\left(
	\begin{array}{ccccc}
		0 & 0 & 0 & 0 & -1 \\
		0 & 0 & 0 & 0 & -1 \\
		0 & 0 & -1 & 0 & 3 \\
		0 & -2 & 3 & 2 & -3 \\
		0 & 1 & -1 & -1 & 1 \\
	\end{array}
	\right)\ ,\ D=\left(
	\begin{array}{ccccc}
		3 & 1 & 2 & -2 & 0 \\
		3 & 3 & 2 & 0 & 0 \\
		-3 & -3 & 0 & 3 & 2 \\
		-5 & -1 & -6 & -1 & 0 \\
		0 & 0 & 1 & 0 & -1 \\
	\end{array}
	\right) \ , \ \vec{\nu}'=\left(
\begin{array}{c}
 -2 \\
 0 \\
 0 \\
 0\\
 0\\
\end{array}
\right)  ~  .
\end{equation*}
Combinatorial flattenings can be obtained using the relations \eqref{flatG} and \eqref{flatAB} and come out to be: 
\begin{equation*}
F_1 = \left(0, 0, 1\right),\,\
F_2 =\left(0, 0, 1\right),\,\
F_3 = \left(0, 3, -2\right),\,\
F_4 = \left(0, 1, 0\right),\,\
F_5 =\left(2, -1, 0\right)  ~.
\end{equation*}
This fixes the following matrices:
\begin{equation}
\vec{f}=\left(
\begin{array}{c}
 0 \\
 0 \\
 0 \\
 0\\
 2\\
\end{array}
\right); \qquad
\vec{f}''=\left(
\begin{array}{c}
 1 \\
 1 \\
 -2 \\
 0\\
 0\\
\end{array}
\right)~.
\end{equation}
\subsubsection{\texorpdfstring{Triangulation of $[\mathbf S^3 \backslash (6_3 \# H)]_{(1,1)}$}{}} \label{ex63}
The manifold $[\mathbf S^3 \backslash (6_3 \# H)]_{(1,1)}$ is obtained by performing the $(1,1)$ Dehn filling on the toroidal boundary associated with the $6_3$ component of the manifold $\mathbf S^3 \backslash (6_3 \# H)$. From the \texttt{SnapPy}, we find that six tetrahedra are required for its ideal triangulation, so $T=\{\Delta_1, \Delta_2,\Delta_3,\Delta_4,\Delta_5,\Delta_6\}$. There are six edges along with the meridian and longitude. The gluing equations can be read off from \texttt{SnapPy} and are given as:
\begin{align}
   z_1' + z_2' + z_3' + z_4 + z_5' &= 2  \pi i \qquad \text{(edge 1)}  \nonumber \\
   z_3 + z_4 + z_4' + z_5' + z_6' + z_6'' &= 2  \pi i \qquad \text{(edge 2)}  \nonumber \\
   z_1 + z_2' + z_2'' + z_3' + z_3'' + z_4'' &= 2  \pi i \qquad \text{(edge 3)}  \nonumber \\
   z_1 + z_3'' + z_4' + z_5 + z_6'' &= 2  \pi i \qquad \text{(edge 4)}  \nonumber \\
   z_1'' + z_2 + z_2'' + z_3 + z_5 + z_5'' + z_6 &= 2 \pi i \qquad \text{(edge 5)}  \nonumber \\
  z_1' + z_1'' + z_2 + z_4'' + z_5'' + z_6 + z_6' &= 2 \pi i \qquad \text{(edge 6)}  \nonumber \\
  2 z_1 - z_1' - z_2'' + 2 z_3'' - z_4 - z_4' - z_4'' - z_5 + z_6 + z_6'' &= 0 \qquad \text{(meridian)}  \nonumber \\
  -z_1 - z_3'' + z_4 + z_5' &= 0 \qquad \text{(longitude)} ~.
   \label{gluingeq63}
\end{align}
The above set of equations is not independent due to the relation between the shape parameters: $z_a+z'_a+z''_a=i \pi$ for $a=1,2,3,4,5,6$. The first seven equations of \eqref{gluingeq63} can be reduced to the following equations:
\begin{align}
    z_4 - z_1 - z_1'' - z_2 - z_2'' - z_3 - z_3'' - z_5- z_5''&=-2 \pi i \nonumber \\
   z_3 - z_4'' - z_5 - z_5'' - z_6&=-i \pi   \nonumber \\
   z_1 + z_4'' - z_2 - z_3&=0    \nonumber \\
    z_1 + z_3'' + z_5 + z_6'' - z_4 - z_4''&=i \pi    \nonumber \\
  z_1'' + z_2 + z_2'' + z_3 + z_5 + z_5'' + z_6 &= 2 \pi i \nonumber \\
  3 z_1 + z_1'' + 2 z_3'' + z_6 + z_6'' - z_2'' - z_5&= 2 \pi i   ~.
\end{align}
This set of equations can be written in a matrix form as $A\vec{z}+ B\vec{z}''=i\pi \vec{\nu}$ from where we can read off the matrices as:
\begin{equation}
    A= \left(
	\begin{array}{cccccc}
		-1 & -1 & -1 & 1 & -1 & 0 \\
		0 & 0 & 1 & 0 & -1 & -1 \\
		1 & -1 & -1 & 0 & 0 & 0 \\
		1 & 0 & 0 & -1 & 1 & 0 \\
		0 & 1 & 1 & 0 & 1 & 1 \\
		3 & 0 & 0 & 0 & -1 & 1 \\
	\end{array}
	\right)\ , \ B= \left(
	\begin{array}{cccccc}
		-1 & -1 & -1 & 0 & -1 & 0 \\
		0 & 0 & 0 & -1 & -1 & 0 \\
		0 & 0 & 0 & 1 & 0 & 0 \\
		0 & 0 & 1 & -1 & 0 & 1 \\
		1 & 1 & 0 & 0 & 1 & 0 \\
		1 & -1 & 2 & 0 & 0 & 1 \\
	\end{array}
	\right) \ , \ 
\vec{\nu}= \left(
\begin{array}{c}
 -2 \\
 -1\\
 0\\
 1\\
 2\\
 2\\
\end{array}
\right) ~.
\end{equation}
Next, we use the longitude equation of \eqref{gluingeq63} which reads as $  z_4 - z_1 - z_3'' - z_5 - z_5''=- i \pi$. This equation along with the constraint \eqref{ABCDSymplectic} can give the matrices $C,D,\vec{\nu}'$ satisfying $C\vec{z}+ D\vec{z}''=i\pi \vec{\nu}'$ and can be written as:
\begin{equation}
    C= \left(
	\begin{array}{cccccc}
		0 & 0 & 0 & 0 & 0 & 0 \\
		0 & 0 & 0 & 0 & 0 & 0 \\
		0 & 0 & 0 & 0 & 0 & 0 \\
		0 & 0 & 0 & -3 & 4 & -1 \\
		0 & 0 & 0 & 0 & 0 & 0 \\
		-1 & 0 & 0 & 1 & -1 & 0 \\
	\end{array}
	\right) \ ,\ D= \left(
	\begin{array}{cccccc}
		-2 & -4 & 2 & -4 & -2 & 4 \\
		0 & -2 & 2 & 0 & 0 & 0 \\
		2 & 0 & 0 & 5 & 3 & -3 \\
		-3 & -3 & 3 & -9 & -2 & 6 \\
		0 & -2 & 2 & 1 & 1 & 1 \\
		0 & 0 & -1 & 0 & -1 & 0 \\
	\end{array}
	\right) \ , \ \vec{\nu}'= \left(
\begin{array}{c}
 -6\\
 2 \\
 0 \\
 -3\\
 -4\\
 -1\\
\end{array}
\right) ~  .
\end{equation}
Combinatorial flattenings can be obtained using the relations \eqref{flatG} and \eqref{flatAB} and come out to be: 
\begin{equation*}
F_1 = \left(0, 0, 1\right),\,\
F_2 =  \left(0, 0, 1\right),\,\
F_3 =  \left(-1, 0, 2\right),\,\
F_4 =  \left(0, 2, -1\right),\,\
F_5 = \left(1, 2, -1\right),\,\
F_6 =  \left(2, 2, -3\right) ~.
\end{equation*}
This fixes the following matrices:
\begin{equation}
\vec{f}=\left(
\begin{array}{c}
 0\\
 0 \\
 -1 \\
 0\\
 1\\
 2\\
\end{array}
\right); \qquad
\vec{f}''=\left(
\begin{array}{c}
 1\\
 1 \\
 2 \\
 -1\\
 -1\\
 -3\\
\end{array}
\right) ~.
\end{equation}

\section{Computing the hyperbolic volume after Dehn filling} \label{appB}
In this section, we compute the generic volume formula of the closed 3-manifold $\hat{M}$ after performing the $(p, q)$ Dehn filling on a manifold $M$ with a single torus boundary. This procedure can also be extended for the case of the manifold with multiple torus boundaries.  We follow the procedure discussed in \cite{NEUMANN}, \cite{aaber2010closed} to perform these computations. To perform the $(p, q)$ Dehn filling on the toroidal boundary of $M$, we choose a basis $(m,\ell) $ of the first Homology group $H_1$ on the torus boundary and paste a solid torus to make $p m +q \ell=1$.
In \cite{NEUMANN}, a changed basis $u=\log m$ and $v=\log \ell$ is used for the convenience of computation.
The parameters $u$ and $v$ are related using a potential function $\Phi$ as:
\begin{equation}\label{uvrelation}
v= \frac{\partial \Phi}{\partial u}.
\end{equation}
The potential function $\Phi$ is an even function of $u$. For the manifolds with a single toroidal boundary, this function is given as \cite{aaber2010closed}:
\begin{equation}\label{phi}
   \Phi(u)=\sum_{a=1}^{\infty}c_{2a} u^{2a} ~,
\end{equation}
where $c_{2i}$ are constants. When we do the $(p,q)$ Dehn filling on the manifold $M$, we get the closed manifold $\hat{M}$. The parameters $u$ and $v$ in such case is related to the integers $p$ and $q$ via the relation:
\begin{equation}\label{uvalue}
    u= \frac{2\pi i}{p+c_2\, q} \quad;\quad v= \frac{1}{2}\frac{\partial \Phi}{\partial u} = \frac{1}{2}\sum_{a=1}^{\infty}2a\, c_{2a}\, u^{2a-1}  ~.
\end{equation} 
After the Dehn filling, the volume will be given as:
\begin{equation}\label{volfor}
\text{Vol}(\hat{M}) = \text{Vol}(M) +\frac{1}{4} \Im\left[u \Bar{v}- \sum_{a=2}^{\infty}(a-1)\, c_{2a}\, u^{2a} \right] ~.
\end{equation}
For a given Dehn filling $(p,q)$, we substitute the values of $u$ and $v$ from \eqref{uvalue} in the \eqref{volfor} to get the volume of $\hat{M}$ as a function of $p$ and $q$.
\subsection{Volume formula for large Dehn filling}
When the Dehn filling is large, i.e. when $p$ or $q$ or both are large integers, we can approximate the volume formula \eqref{volfor} by its leading term which is a quadratic term in $p$ and $q$. In such case, we will have:
\begin{align}\label{volforLarge}
\text{Vol}(\hat{M}) &= \text{Vol}(M) +\frac{1}{4} \Im\left[\frac{4 \pi^2 \bar{c}_2}{(p+c_2\,q)(p+\bar{c}_2\,q)} \right] + \mathcal{O}\left[\frac{1}{(p^4, q^4)}\right] + \ldots \nonumber\\
&= \text{Vol}(M) - \frac{ \pi^2 \Im[c_2]}{p^2 + \abs{c_2}^2 q^2 + 2pq \Re[c_2]} +  \mathcal{O}\left[\frac{1}{(p^4, q^4)}\right] + \ldots ~.
\end{align}
The constant $c_2$ appearing in this formula contains the information of the modulus of the cusp \cite{aaber2010closed} of the manifold $M$. In the following, we will obtain the volume of the closed 3-manifold obtained by the  $(p,q)$ Dehn filling of $[\mathbf S^3 \backslash (\mathcal{K} \# H)]_{(1,1)}$ using the above formula. The value of the constant $c_2$ for the manifold $[\mathbf S^3 \backslash (\mathcal{K} \# H)]_{(1,1)}$ can be computed using \texttt{SnapPy} by using the following Python $3.0$ code:
\begin{python}
    import snappy
    (*Step 1: We start with defining a Manifold M with 
 two torus boundary. *)
    M=snappy.Manifold()
    (*Step 2:  We perform a (p, q) Dehn filling on the torus boundary  corresponding to the knot component*)
    M.dehn_fill((p,q),0)
    (*Step 3:  We do a filled triangulation of the manifold and get a manifold with a single torus boundary.*)
    MF=M.filled_triangulation()
    (*Step 4: We obtain the information of the modulus of the cusp*)
    MF.cusp_info(`modulus')
\end{python}
\textbf{$\bullet$ Volume formula for $(p,q)$ Dehn filling of $[\mathbf S^3 \backslash (4_1 \# H)]_{(1,1)}$  :} \\
Here $M=[\mathbf S^3 \backslash (4_1 \# H)]_{(1,1)}$ and $\hat{M}=[\mathbf S^3 \backslash (4_1 \# H)]_{(1,1),(p,q)}$ where we have already discussed these notations in section \ref{sec3}. The value of $c_2$ for the manifold $M$ comes out to be 
\begin{equation}
    c_2 = 3.464101615137755 \, i ~.
\end{equation}
Thus, for large values of Dehn filling, we will have:
\begin{equation}\label{volfor41}
   \text{Vol}(\hat{M}) = \text{Vol}(M)-\frac{{34.1893}}{p^2+ 12q^2}+O\left[\frac{1}{(p^4, q^4)}\right]+....   ~. 
\end{equation}
\textbf{$\bullet$ Volume formula for $(p,q)$ Dehn filling of $[\mathbf S^3 \backslash (5_2 \# H)]_{(1,1)}$  :} \\
Here $M=[\mathbf S^3 \backslash (5_2 \# H)]_{(1,1)}$ and $\hat{M}=[\mathbf S^3 \backslash (5_2 \# H)]_{(1,1),(p,q)}$ where we have already discussed these notations in section \ref{sec3}. The value of $c_2$ for the manifold $M$ comes out to be 
\begin{equation}
    c_2 = 0.490244667506610 + 2.979447066478973\, i ~.
\end{equation}
Thus, for large values of Dehn filling, we will have:
\begin{equation}\label{volfor52}
   \text{Vol}(\hat{M}) = \text{Vol}(M)- \frac{29.406}{ p^2+0.980489 p q+9.11744 q^2}+ \mathcal{O}\left[\frac{1}{(p^4, q^4)}\right]+\ldots ~.
\end{equation}
\textbf{$\bullet$ Volume formula for $(p,q)$ Dehn filling of $[\mathbf S^3 \backslash (6_1 \# H)]_{(1,1)}$  :} \\
Here $M=[\mathbf S^3 \backslash (6_1 \# H)]_{(1,1)}$ and $\hat{M}=[\mathbf S^3 \backslash (6_1 \# H)]_{(1,1),(p,q)}$ where we have already discussed these notations in section \ref{sec3}. The value of $c_2$ for the manifold $M$ comes out to be 
\begin{equation}
    c_2 = 0.173261721741264 + 2.564798632228027 \, i ~.
\end{equation}
Thus, for large values of Dehn filling, we will have:
\begin{equation}\label{volfor61}
   \text{Vol}(\hat{M}) = \text{Vol}(M) -\frac{25.3135 }{ p^2+0.346523 p q+6.60821 q^2} +\mathcal{O}\left[\frac{1}{(p^4, q^4)}\right]+ \ldots ~.
\end{equation}
\textbf{$\bullet$ Volume formula for $(p,q)$ Dehn filling of $[\mathbf S^3 \backslash (6_2 \# H)]_{(1,1)}$  :} \\
Here $M=[\mathbf S^3 \backslash (6_2 \# H)]_{(1,1)}$ and $\hat{M}=[\mathbf S^3 \backslash (6_2 \# H)]_{(1,1),(p,q)}$ where we have already discussed these notations in section \ref{sec3}. The value of $c_2$ for the manifold $M$ comes out to be 
\begin{equation}
    c_2 = -0.255686381582274 + 3.498585842104345 \, i ~.
\end{equation}
Thus, for large values of Dehn filling, we will have:
\begin{equation}\label{volfor62}
   \text{Vol}(\hat{M}) = \text{Vol}(M) -\frac{{34.5297 i}}{ p^2-0.511373 p q+12.3055 q^2} +\mathcal{O}\left[\frac{1}{(p^4, q^4)}\right]+\ldots~.
\end{equation}
\textbf{$\bullet$ Volume formula for $(p,q)$ Dehn filling of $[\mathbf S^3 \backslash (6_3 \# H)]_{(1,1)}$  :} \\
Here $M=[\mathbf S^3 \backslash (6_3 \# H)]_{(1,1)}$ and $\hat{M}=[\mathbf S^3 \backslash (6_3 \# H)]_{(1,1),(p,q)}$ where we have already discussed these notations in section \ref{sec3}. The value of $c_2$ for the manifold $M$ comes out to be 
\begin{equation}
    c_2 = 5.510570258265165 \, i ~.
\end{equation}
Thus, for large values of Dehn filling, we will have:
\begin{equation}\label{volfor63}
   \text{Vol}(\hat{M}) = \text{Vol}(M) -\frac{54.3871 }{ p^2+30.3664 q^2}+\mathcal{O}\left[\frac{1}{(p^4, q^4)}\right]+ \ldots~.
\end{equation}

\bibliographystyle{unsrt}
\bibliography{refdehn.bib}

\begin{thebibliography}{10}

\bibitem{witten1989}
Edward Witten.
\newblock Quantum field theory and the jones polynomial.
\newblock {\em Communications in Mathematical Physics}, 121(3):351--399, 1989.

\bibitem{lab1999}
JMF Labastida.
\newblock Chern-simons gauge theory: Ten years after.
\newblock In {\em AIP Conference Proceedings}, volume 484, pages 1--40.
  American Institute of Physics, 1999.

\bibitem{heinonen1998}
Olle Heinonen.
\newblock {\em Composite fermions: a unified view of the quantum Hall regime}.
\newblock World Scientific, 1998.

\bibitem{murthy2003}
Ganpathy Murthy and R~Shankar.
\newblock Hamiltonian theories of the fractional quantum hall effect.
\newblock {\em Reviews of Modern Physics}, 75(4):1101, 2003.

\bibitem{marino2005}
Marcos Marino.
\newblock Chern-simons theory and topological strings.
\newblock {\em Reviews of Modern Physics}, 77(2):675, 2005.

\bibitem{witten1991}
Edward Witten.
\newblock Quantization of chern-simons gauge theory with complex gauge group.
\newblock {\em Communications in Mathematical Physics}, 137(1):29--66, 1991.

\bibitem{gukov2005}
Sergei Gukov.
\newblock Three-dimensional quantum gravity, chern-simons theory, and the
  a-polynomial.
\newblock {\em Communications in mathematical physics}, 255:577--627, 2005.

\bibitem{Balasubramanian:2018por}
Vijay Balasubramanian, Matthew DeCross, Jackson Fliss, Arjun Kar, Robert~G.
  Leigh, and Onkar Parrikar.
\newblock {Entanglement Entropy and the Colored Jones Polynomial}.
\newblock {\em JHEP}, 05:038, 2018.

\bibitem{Dimofte:2016pua}
Tudor Dimofte.
\newblock {Perturbative and nonperturbative aspects of complex
  Chern\textendash{}Simons theory}.
\newblock {\em J. Phys. A}, 50(44):443009, 2017.

\bibitem{Duan:2022ryd}
Zhihao Duan and Jie Gu.
\newblock {Resurgence in complex Chern-Simons theory at generic levels}.
\newblock {\em JHEP}, 05:086, 2023.

\bibitem{Freed:2022yae}
Daniel~S. Freed and Andrew Neitzke.
\newblock {3d spectral networks and classical Chern-Simons theory}.
\newblock 8 2022.

\bibitem{Garoufalidis:2023ipa}
Stavros Garoufalidis, Matthias Storzer, and Campbell Wheeler.
\newblock {Perturbative invariants of cusped hyperbolic 3-manifolds}.
\newblock 5 2023.

\bibitem{ad}
Aditya Dwivedi, Siddharth Dwivedi, Bhabani~Prasad Mandal, Pichai Ramadevi, and
  Vivek~Kumar Singh.
\newblock Topological entanglement and hyperbolic volume.
\newblock {\em Journal of High Energy Physics}, 2021(10):1--37, 2021.

\bibitem{gang}
Dongmin Gang, Mauricio Romo, and Masahito Yamazaki.
\newblock All-order volume conjecture for closed 3-manifolds from complex
  chern--simons theory.
\newblock {\em Communications in Mathematical Physics}, 359(3):915--936, 2018.

\bibitem{hikami1}
Kazuhiro Hikami.
\newblock Hyperbolic structure arising from a knot invariant.
\newblock {\em International Journal of Modern Physics A}, 16(19):3309--3333,
  2001.

\bibitem{hikami2}
Kazuhiro Hikami.
\newblock Generalized volume conjecture and the a-polynomials: the
  neumann--zagier potential function as a classical limit of the partition
  function.
\newblock {\em Journal of Geometry and Physics}, 57(9):1895--1940, 2007.

\bibitem{dimofte2013}
Tudor Dimofte.
\newblock Quantum riemann surfaces in chern-simons theory.
\newblock {\em Advances in Theoretical and Mathematical Physics},
  17(3):479--599, 2013.

\bibitem{dimofte20131}
Tudor Dimofte and Stavros Garoufalidis.
\newblock The quantum content of the gluing equations.
\newblock {\em Geometry \& Topology}, 17(3):1253--1315, 2013.

\bibitem{faddeev}
Ludwig~D Faddeev and Rinat~M Kashaev.
\newblock Quantum dilogarithm.
\newblock {\em Modern Physics Letters A}, 9(05):427--434, 1994.

\bibitem{SnapPy}
Marc Culler, Nathan~M. Dunfield, Matthias Goerner, and Jeffrey~R. Weeks.
\newblock Snap{P}y, a computer program for studying the geometry and topology
  of $3$-manifolds.
\newblock Available at \url{http://snappy.computop.org}.

\bibitem{Bae}
Jin-Beom Bae, Dongmin Gang, and Jaehoon Lee.
\newblock 3d n= 2 minimal scfts from wrapped m5-branes.
\newblock {\em Journal of High Energy Physics}, 2017(8):1--23, 2017.

\bibitem{NEUMANN}
Walter~D. Neumann and Don Zagier.
\newblock Volumes of hyperbolic three-manifolds.
\newblock {\em Topology}, 24(3):307--332, 1985.

\bibitem{mathews2022symplectic}
Daniel~V Mathews and Jessica~S Purcell.
\newblock A symplectic basis for 3-manifold triangulations.
\newblock {\em arXiv preprint arXiv:2208.06969}, 2022.

\bibitem{Neumann1992}
Walter~D. Neumann.
\newblock Combinatorics of triangulations and the chern-simons invariant for
  hyperbolic 3-manifolds.
\newblock 1992.

\bibitem{aaber2010closed}
John~W Aaber and Nathan Dunfield.
\newblock Closed surface bundles of least volume.
\newblock {\em Algebraic \& Geometric Topology}, 10(4):2315--2342, 2010.

\end{thebibliography}


\end{document}